\documentclass[letterpaper,  singlespace,  twocolumn]{emulateapj}

\usepackage{natbib} 
\usepackage{graphicx}
\usepackage{bm}
\usepackage{hyperref}
\usepackage{subfigure}
\usepackage{amsmath}
\usepackage{amssymb}
\usepackage{wasysym}
\usepackage{verbatim}

\makeatletter
\def\cutinhead#1{\noalign {\vskip 0ex}\hline \@ptabularcr \noalign {\vskip 0.5ex}\multicolumn {\LT@cols }{c}{#1}\@ptabularcr \noalign {\vskip .8ex}\hline \@ptabularcr \noalign {\vskip 0.2ex}}
\def\tablehead#1{\@table@not@headedfalse \kill \caption {\\\@tablecaption }\\ \hline \hline
 \\[0.2ex] #1\hskip \tabcolsep \\[.7ex] \hline \\[-0.25ex] \endfirsthead \caption []{--- \emph {Continued}}\\ \hline \hline \\[-1.7ex] #1\hskip \tabcolsep \\[.7ex] \hline \\[-1.5ex] \endhead \hline \endfoot}
\makeatother

\def\FeH{[\mathrm{Fe}/\mathrm{H}]}
\def\alphaFe{[\alpha/\mathrm{Fe}]}

\begin{document}

\title[Not All Stars Are the Sun: Empirical Calibration of the Mixing Length for Metal-Poor Stars Using One-dimensional Stellar Evolution Models]
{Not All Stars Are the Sun: Empirical Calibration of the Mixing Length for Metal-Poor Stars Using One-dimensional Stellar Evolution Models}

\author{M. Joyce\altaffilmark{1}\altaffilmark{2} and B. Chaboyer\altaffilmark{1}}
\affil{$^{1}$Department of Physics and Astronomy,  Dartmouth College,  Hanover,  NH 03755}
\affil{$^{2}$South African Astronomical Observatory, PO Box 9, Observatory, Cape Town 7935, South Africa}

\begin{abstract}
Theoretical stellar evolution models are constructed and tailored to the best known, observationally derived characteristics of metal-poor ([Fe/H]$\sim-2.3$) stars representing a range of evolutionary phases: subgiant HD140283, globular cluster M92, and four single, main sequence stars with well-determined parallaxes: HIP46120, HIP54639,  HIP106924, and WOLF1137. It is found that the use of a solar-calibrated value of the mixing length parameter $\alpha_{\text{MLT}}$ in models of these objects is ineffective at reproducing their observed properties. Empirically calibrated values of $\alpha_{\text{MLT}}$ are presented for each object, accounting for uncertainties in the input physics employed in the models. It is advocated that the implementation of an adaptive mixing length is necessary in order for stellar evolution models to maintain fidelity in the era of high precision observations. 
\end{abstract}


\maketitle
%----------------------------------------- section 1: Introduction --------------------------------------------
\section{Introduction}
Characterizing heat transport in stars is notoriously complicated, and the task of reproducing the physics involved with high precision on stellar evolutionary timescales is a long-standing problem in stellar modeling.
For this reason, convection in 1-D stellar evolution codes is addressed primarily through a framework known as mixing length theory (MLT). 
Impeded by computational limitations, stellar codes invoke MLT to parameterize a non-linear, turbulent, 3-dimensional process in a 1-dimensional manner.

A simplified picture of convection was developed by Ludwig \cite{LP}, whose framework invoked analogy with molecular heat transfer. Conceiving of the bulk transfer of fluids as taking place in discrete ``parcels,''  we may define a region inside of a convective area and map the vertical trajectory of a particular parcel with uniform physical characteristics. Assuming the parcel is in pressure, but not thermal, equilibrium with its surroundings, a hot parcel of material rises {upward} to a cooler region, which causes it to expand and denature; analogously, cooler parcels fall and compress. 
The ``mixing length'' then refers to the characteristic vertical distance, or mean free path, that such a parcel can travel before losing its definition. The mixing length is defined in units of pressure scale height, $d \ln(P)/d \ln(T)$. This framework was first applied to stellar interiors by Erika B\"{o}hm-Vitense \citep{EBV}, whose seminal paper on solar convection has been used to guide stellar models for over half a century.

Much progress has been made since to handle convection in a more sophisticated way. Convective overshoot, for example, is also invoked in 1-D codes in the form of an adjustable parameter describing the degree of permeability of the convective boundary layer.  With the advent of supercomputing, 3-D hydrodynamical models are available, and they are able to characterize convection much more authentically than any lower-dimensional formulation { (see, e.g.\ \citet{Nord82}, \citet{Deupree85}, \citet{Freytag96}, \citet{Ludwig99}, \citet{Arnett09}, \citet{Trampedach}; many others)}. However, they can only do so on { very short timescales, relative to evolutionary time}, and typically only for a region near the surface of the star. 

While the state of the art for modeling atmospheres and surface convection is quite advanced, it is another undertaking entirely to address mixing across evolutionary time and over the range of temperatures and densities { within a} stellar interior. There have been few attempts to generalize MLT to higher dimensions, but these and other improvements to the framework have been examined. Such investigations include {
(1) non-local versions of MLT \citep{NLMLT,Grossman93}; 
(2) those which invoke full spectrum turbulence (e.g.\ \citet{Canuto91});
(3) calibrations of the mixing length parameter against 3-D radiative hydrodynamical simulations (e.g. \citet{Ludwig99, Trampedach, MagicMLT, 321D});
and
(4) the development of models which seek to remove mixing length theory's approximations and which are based directly on 3-D simulations (e.g. \citet{Lydon92,321D}).} 
Despite these efforts, MLT endures most often in a form not hugely different from that in which it was first conceived.

Some obvious shortcomings of this theory follow from the na{\"i}ve physical assumptions it supposes. {In particular, mixing length theory ignores convective overshooting, despite the fact that the boundaries of the convective region(s) in a stellar interior are not physically rigid. Many other issues are caused especially by the assumption of strictly vertical paths. MLT supposes a only
{a simple, 1-D trajectory for a parcel; real convection, on the other hand, involves the continuous shearing, fragmenting, reorientation, and deletion of the flow channels through which material travels.}

Likewise, MLT incorrectly assumes symmetry between upflows and downflows. In reality, there is no symmetry between channels that carry material towards the surface of the star and channels that carry it away. Plasma flowing upward expands as its density drops, creating broad convection cells surrounded by an interconnected network of thinner downflow lanes. Since the upflows are expanding as they move material towards lower-density layers of the star, turbulence in the upward lanes will be smoothed out. In the downflow lanes, the material is traveling against the density gradient, which enhances turbulence. Also, due to the density gradient, conservation laws dictate that upflows must donate mass to the downflows \citep{SteinNord89}.

The asymmetry between up- and downflows in real convective regions means that most of the plasma in the upflows originates from the deep interior of the convection zone, making the upflows isentropic \citep{SteinNord89}. The downflows, however, contain contributions from over-turning at every point on the way. Some of this material is from the surface,  where it underoes radiative cooling and lost entropy. The downflows therefore incorporate a large range of low entropies and are hence denser than the upflows. This facilitates the downflows' penetration into denser interior regions, causing turbulence.  Mass conservation at each layer, combined with momentum conservation and the fact that the downflows are slightly denser, means that (1) downflow speeds are higher and (2) downflows occupy a smaller area than the upflows. These conditions in combination result in ``negative kinetic flux,'' or the inward transfer of kinetic energy \citep{SteinNord89}. Such plasma physics is well understood, but MLT cannot account for these asymmetries.

Lastly, radiative hydrodynamics has shown that the idea of a well-defined convective bubble is not valid; { there is no quantized unit of convection in stellar interiors, but rather a continuum of upflows and cooler, turbulent downflows. }

In addition to the above issues, the mixing length 
{parameterizes the entropy jump between the top of the convective envelope and the asymptotically adiabatic portion of the convection zone. Neither this entropy jump, nor the depth of the convective envelope which it implies, are observable.} Given its lack of a priori physical justification, $\alpha_{\text{MLT}}$ must be empirically determined. In stellar codes, $\alpha_{\text{MLT}}$ is varied in solar reproductions until the model star's temperature, luminosity, and surface metal abundance at the solar age reflect the observed solar values to their best known accuracy. Naturally, { empirical data} are known to much greater precision and with much greater certainty for the Sun than for any other star.

In the same way that $\alpha_{\text{MLT}}$ is not a physical constant, it is also not a computational one. Because it is a free parameter, the value of $\alpha_{\text{MLT}}$ must be determined on an individual basis in { each stellar evolution code,} and its value as determined by different codes will reflect the prescriptive differences among those codes---even though the target features to be reproduced are identical.

Most importantly, not all stars are the Sun. While it is well understood that our nearest star is not a valid representation of stars in general, it remains the standard in stellar evolution codes to calibrate the mixing length according to solar specifications---uniquely---and then apply this $\alpha_{\text{MLT}}$ value to stellar models of a wide range of masses and compositions.

However, there has been growing evidence that the use of a solar-calibrated mixing length is not always appropriate. Attempts at modeling the $\alpha$-Centauri binary system (which consists of $1.1$ and $0.9\, M_\odot$ stars with nearly solar compositions) have suggested, for many years, that a solar-calibrated mixing length may not be suitable for all stars (e.g.\ \citet{GD2000}). Stellar models of the near-solar metallicity 61 Cygni binary system ($M = 0.7$ and $0.6\,M_\odot$) have suggested that a subsolar mixing length should be used for lower-mass stars \citep{Kervella08}. Three-dimensional radiative hydrodynamic simulations of convection in the surface layers of stars
{predict that the mixing length used in stellar evolution codes should depend on the composition and surface gravity of the star (e.g.\ \citet{Freytag99}; \citet{Trampedach}; \citet{MagicMLT}).} We note, however, that such studies do not predict significantly different effective temperatures for stellar models than what is found when using a solar-calibrated mixing length \citep{SC2015}}.

Finally, asteroseismology of \textit{Kepler} target stars has found that the observed radii and effective temperatures of red giant stars require a variable mixing length (e.g.\ \citet{Bonaca}; \citet{Tayar}), with the most recent work suggesting that a metallicity-dependent mixing length requiring a decrease of order $\Delta \alpha \sim 0.2\ $ per dex should be used for red giant stars.

It has also been demonstrated explicitly that a solar-calibrated mixing length is ineffective at modeling stellar conditions that deviate significantly from the Sun's. In particular, \cite{Creevey} used interferometry to resolve the metal-poor subgiant HD 140283 and determine its radius.  They found that the solar-calibrated value of $\alpha_{\text{MLT}}$ does not produce models that match the observed temperature and luminosity of the highly metal-poor star ($\FeH = -2.2$) HD 140283. We corroborate this result and also find that lower values of $\alpha_{\text{MLT}}$ are necessary to fit other stellar objects with metallicities near HD 140283's ([Fe/H]$=-2.2$ to $-2.4$). Our findings are consistent with results calibrated against data from NASA's {\it Kepler} mission \citep{Bonaca}, which suggest that the mixing length needs to vary with stellar properties.

Nevertheless, mixing length theory is still implemented in {most} stellar evolution codes. This includes
the 1-D stellar evolution code DSEP: the Dartmouth Stellar Evolution Program \citep{Dotter}), which is used to generate all stellar models in this study. The standard will persist in the foreseeable future; however, we aim to mitigate, in part, MLT's biasing impact on stellar evolution calculations through considering empirical calibrations to 
{stars which differ significantly from the Sun.}
 With the advent of highly accurate observational data on metal-poor stars throughout their evolution, stellar interior environments that deviate significantly from solar conditions provide laboratories in which we can test the extent of the solar mixing length's usefulness.

This paper is laid out as follows. First, we describe the procedure for calibrating $\alpha_{\text{MLT}}$ and present a series of solar-calibrated mixing length values for a range of physical prescriptions within DSEP. Next, we present the mass-$\alpha_{\text{MLT}}$ combinations of DSEP stellar tracks which fit the observed properties of HD 140283, as determined by \citet{Creevey}. We also elect to examine the effects of uncertainties in the input physics on the stellar models by considering modifications to the atmospheric boundary conditions and the efficiency of diffusion (hereafter denoted with the parameter $\eta_{\text{D}}$).
Although these are just two of many sources of uncertainty (others include, for instance, opacities and nuclear reaction rates), we choose to focus on surface boundary conditions and diffusion because they have been shown in, e.g.\ \citet{Chab95}, \citet{Chab98}, and \citet{ChabKrauss} to be the more significant contributers of uncertainty in stellar models.

We next present best-fitting isochrones to the metal-poor globular cluster M92, which, due to its compositional similarity, serves as a population analogous to HD 140283, and likewise allows us to consider the impact of variable mixing length on the red giant branch. 
Following this, we present empirical, best-fit mixing length values for four single, metal-poor, main sequence 
{ stars (hereafter referred to as ``subdwarfs,'' as they are fainter at a given color compared to solar-metallicity main sequence stars)} which have {metallicities similar} to those of M92 and HD 140283. These stars are HIP46120, HIP54639, HIP106924, and WOLF1137. We conclude with the argument that our results demonstrate (1) a subsolar trend for $\alpha_{\text{MLT}}$ in metal-poor stars and (2) the necessity of an { adaptive treatment} of the mixing length.

%---------------------------------------------------------------------------------
\section{Solar Mixing Length Calibration}
One of many issues with the MLT framework, and free parameters in general, is that the value of $\alpha_{\text{MLT}}$ is subject to its computational environment; it ``absorbs'' computational and physical artifacts unique to the code in which it is implemented. Because the physics used within stellar evolution codes varies from model to model, $\alpha_{\text{MLT}}$ must be calibrated according to each code. Likewise, it must be recalibrated whenever the physical prescription within a code is altered.

We calibrate a solar model by adjusting the mixing length, initial helium abundance (Y), and initial heavy element abundance (Z) (which affect the chemical evolution degenerately) until the model reproduces the observed solar radius, luminosity, and surface $Z$/$X$ abundance at the solar age to better than 0.1\% accuracy. We solve the stellar structure equations to a tolerance of one part in $10^5$, and find that we can compute the mixing length to five digit accuracy before adjacent input values give redundant residuals in $\log{R_{\text{model}}}/\log{R_{\odot}}$ and $\log{L_{\text{model}}}/\log{L_{\odot}}$. The mixing length obtained through this minimization is $\alpha_{\odot, \text{DSEP}}$.

In typical circumstances, this mixing length value becomes a constant within the code, assumed in subsequent { calculations for models of all types of stars.} In this study, we conduct a series of solar mixing length calibrations under different prescriptions for DSEP's physics. These calibrations demonstrate (1) the degree to which $\alpha_{\text{MLT}}$ is affected by other components of the code, and (2) the differences within the stellar interior implied by various physical models.  

%%%%%%%%%%%%%%%%%%%%%%%%% NEW SECTION ON DIFFUSION %%%%%%%%%%%%%%%%%%%%%%%%%%%%%%%%%%%%%
{ We provide solar mixing length calibrations for four different prescriptions of the modeling physics. We elect to vary the surface boundary conditions and diffusive efficiency because there is considerable uncertainty associated with each \citep{Chab95}, and both of these parameters can have a significant impact on the predicted effective temperature of the stellar models (see discussion in \citet{Trampedach} and \citet{SC2015}). We explore two choices of outer boundary conditions: the PHOENIX model atmospheres, which are standard in DSEP, and the Eddington approximation to the grey atmosphere \citep{Eddingtonttau}. We also vary the effectiveness of diffusion, which is enhanced or suppressed by changing the parameter $\eta_{\text{D}}$. The standard diffusive ``strength'' is represented by a value of $\eta_{\text{D}}=1.0$.

The treatment of diffusion in DSEP
{follows the prescription of \citet{Thoul94} and}
 includes thermal diffusion and gravitational settling, but not radiative levitation.
{In this prescription, H, He, and heavy elements are diffused, where heavy elements are represented as a single species assumed to diffuse at the same rate as fully ionized iron \citep{Chab2001}. A thorough discussion of DSEP's formalism for diffusion, the uncertainties therein, and how these processes contribute to DSEP's $\eta_{\text{D}}$ parameter are described in \citet{Chab2001}).}
For our purposes, it is sufficient to consider $\eta_{\text{D}}$ to be a parameterization of the coefficients embedded in the equations governing thermal diffusion and gravitational settling, and to be a parameter which expresses a measure of how quickly the diffusive process proceeds globally. Two of the physical configurations investigated in this study use altered diffusion: one with diffusion suppressed to half its default efficiency ($\eta_{\text{D}}=0.5$), and one enhanced ($\eta_{\text{D}}=1.5$).

The atmospheric boundary conditions used by DSEP can be specified from a number of possibilities. PHOENIX model atmospheres \citep{Hauschildt} are used by default to determine the surface boundary conditions for temperatures up to 10,000 K and $\log g=5.5$ (all of the models produced in this study are well beneath these conditions). DSEP does the matching between the interior solution and the surface boundary conditions at an optical depth of $\tau = 2/3$. 
{This approach has been adopted as it leads to stellar models which are in broad agreement with the observations (e.g.\  \citet{Dotter07} , \citet{Dotter}).}
More details on DSEP's handling of both diffusion and boundary conditions, and justifications for choices regarding these implementations, can be found in \citet{Dotter}.

Table \ref{solar} shows the value of the solar-calibrated mixing length for four physical configurations: The first (referred to henceforth as our ``default'' or ``canonical'' configuration) uses a PHOENIX model atmospheres as boundary conditions and DSEP's default implementation of diffusion. The second uses a grey model atmosphere and the same diffusive efficiency. The latter two revert back to the PHOENIX model atmospheres, but suppress and enhance diffusion, respectively.}

Due to turbulence, the convective regions of stars are well-mixed compared to their radiative counterparts, and mixing is assumed to be instantaneous compared to evolutionary timescales. Because the mixing length is defined to be a mean free path for fluid parcels, it is, in some sense, a proxy for the mixing efficiency within the convective region. Speaking strictly in the language of MLT, a larger mixing length indicates that a parcel is traveling further in space before 
denaturing, thereby representing a more efficient exchange of heat from different points along {its trajectory.} Larger mixing lengths lead to more efficient transport of heat within the star, which in turn yields small convection zones. Hence, the star becomes more compact and displays a {higher} surface temperature.

Despite the fact that the MLT framework and the physical prescription for diffusion apply to entirely separate regions of the star (convective and radiative, respectively), {changes in mixing length can mimic the effect on global parameters of modifying the efficiency of diffusion.}

{The physical consequences of increasing the efficiency of diffusion include a decrease in hydrogen in the center of the model, which leads to shorter main sequence lifetimes, and an increase in hydrogen abundance in the outer region of the star, which causes an increase in opacity in the surface layers. This increase in opacity leads to a decrease in the surface temperature of the model. Raising the diffusive efficiency ($\eta_D$) and lowering the mixing length therefore both lead to a decrease in the model star's effective temperature. 
We emphasize, however, that changes in the mixing length and in the efficiency of diffusion are not physically related, but merely manifest the same way in terms of surface observeables.

Obtaining lower and higher values in our solar calibrations of $\alpha_{\text{MLT}, \odot}$---where the effective temperature is fixed at the solar value---for models with suppressed ($\eta_{\text{D}}=0.5$) and enhanced ($\eta_{\text{D}}=1.5$) diffusion, respectively, is therefore consistent with our expectations. }

\begin{table} 
\centering 
\caption{Solar-Calibrated Mixing Length Values for Various Physical Configurations}
\begin{tabular}{ l l l   r   l   }  
\hline\hline
 Atmosphere & $\eta_{\text{D}}$ & $\alpha_{\odot}$ & $Y_{\text{in}}$ & $Z_{0}$ 
\\ \hline
PHOENIX & 1.0&	 1.9258  & 0.275 & 0.019 \\
Grey 	& 1.0&	 1.8205 &  0.282 & 0.019\\
PHOENIX & 0.5& 	 1.8292 &   0.277 &  0.0176 \\
PHOENIX & 1.5&	 1.9780 &  0.282 & 0.0192 \\
 \hline
\end{tabular}
\tablecomments{ The solar-calibrated mixing length $\alpha_{\text{MLT, DSEP}}$ is computed for {models} with varying diffusive efficiency and atmospheric boundary conditions. In all cases, the solar luminosity and radius at the solar age were reproduced to within 0.1\% of the true solar values. The surface abundance Z/X was reproduced at the solar age to within 1\% of the true solar value. 
}
\label{solar}
\end{table}

\section{Stellar Models}
\label{stellar}
A thorough discussion of the DSEP code can be found in \citet{Chab06} and \citet{Dotter}. Major adjustments within the code since 2008 are discussed in \citet{me}, and include updates to nuclear reaction rates \citep{Adel, Marta}.

We use DSEP to generate grids of stellar tracks which are then interpolated to construct isochrones. 
In the case of comparing DSEP's predictions to HD 140283, we source our data directly from the stellar tracks.

To fit HD 140283, we probe a parameter space encompassing masses of 0.7--0.85 M$_{\odot}$ and mixing lengths ranging from $\alpha_{\text{MLT}}=0.5$ to $2.0$ by generating a grid of stellar tracks in mass increments of 0.01 M$_{\odot}$ and $\alpha_{\text{MLT}}$ increments of 0.1. To fit M92 and the four main sequence stars, we generate isochrones in mixing length increments of roughly 0.05---the sampling is not constant across the entire mixing length spectrum due to the fact that changes in the mixing length at low values of $\alpha_{\text{MLT}}$ have a much greater impact on the isochrone's effective temperature than do identical changes at higher $\alpha_{\text{MLT}}$.

Each isochrone involves the generation of two grids of stellar tracks. A high-mass grid is generated using DSEP's standard, analytical equation of state, which includes Coulomb corrections \citep{ChabKim95}, and encompasses tracks of mass 0.65 to 1.0 M$_{\odot}$, in  increments of roughly 0.03 M$_{\odot}$. A separate grid of tracks is generated using the more sophisticated (and more computationally demanding) FREE EOS equation of state \citep{FreeEOS}, for $M<0.65M_{\odot}$. FREE EOS is invoked only in the low-mass regime, where it has a significant impact on the tracks.

{We compute isochrones for ages 8 to 16 Gyr, in increments of 1 Gyr. 
To account for uncertainty in the ages of the objects, we consider all of these as potential fits to M92 and the four main sequence stars. However, only isochrones within the range of 11-15 Gyr match the observations, and we therefore concentrate on analyzing this range.}

\section{Fits To HD 140283}
HD 140283 is significant in both its metal-poorness and proximity. Thanks to the latter, it is the only very metal-poor star whose radius we know empirically from interferometry.

{ In their recent paper, \citet{Creevey} (hereafter C15) thoroughly constrain the observational features of HD 140283. Over 4 nights in 2012 and 2014, C15 used CHARA array interferometric observations to determine the angular diameter of HD 140283.} Using these data, { they derive a (limb-darkened) radius }of $R=2.21 \pm 0.08 R_{\odot}$, a luminosity of $4.12 \pm 0.10 L_{\odot}$, and $T_{\text{eff}}=5543$, assuming an interstellar extinction of $A_V = 0.0$ mag. However, $A_V$ is uncertain, and in adopting a maximal error bar of $A_V=0.1$, the constraints on luminosity extend up to $4.47 L_{\odot}$, and on $T_{\text{eff}}$ to $5647$ K. These values provide the bounds on the HR diagram to which we fit. 
{ As in C15, we adopt $\FeH= -2.46 \pm 0.014$ based on the PASTEL catalogue \citep{PASTEL}. }

In deriving the star's model-dependent properties---mass, age, initial metal abundance---C15 employ the CESAM stellar evolution code. This code invokes libraries and formalisms that are standard among stellar evolution codes, including OPAL opacities \citep{Ig}, the NACRE nuclear reaction rates, MARCS \citep{MARCS} model atmospheres, and the MLT framework for modeling convective regions. Using { this input physics}, they create a set of models spanning a range of masses, initial metallicities, and mixing lengths to find the combination that best reproduces derived luminosity, effective temperature, and present metallicity of HD 140283.

Irrespective of the mixing length value, C15 use stellar tracks to determine that the mass of HD 140283 falls between 0.75 and 0.84 M$_{\odot}$ based on the observed luminosity, { the physically reasonable assumption that the star is not older than the age of the universe ($t<14$ Gyr)}, and the primordial helium abundance ($Y_i=0.245$). Having inferred these mass constraints without co-optimizing the mixing length, they are { then left with only zero-age metallicity $Z_{0}$---the primordial metal abundance by weight specified in the pre-MS model---and $\alpha_{\text{MLT}}$ as parameters that cannot be constrained directly by observations.}

In adjusting both $Z_{0}$ and $\alpha_{\text{MLT}}$, C15 find, firstly, that no model with $\alpha_{\text{MLT}} = \alpha_{\odot}$ (in the CESAM code, $\alpha_{\odot}= 2.0$) can reproduce the effective temperature $T_{\text{eff}}$ and interferometrically derived radius $R$ of HD 140283. Rather, they find that a mixing length half this value ($\alpha_{\text{MLT}}=1.0$) is needed. Considering the smallest possible mass (0.78 M$_{\odot}$---see Section 5 for a discussion of the mass--mixing length degeneracy) within their constraints, they can push $\alpha_{\text{MLT}}$ up only as far as 1.5, or $0.75$ $\alpha_{\odot}$. Across the full mass spectrum, the best-fitting mixing length drops as low as $0.45 \alpha_{\odot}$ ($\alpha_{\text{MLT}}=0.9$) at 0.82 M$_{\odot}$.

Their Figure 7 shows a set of tracks with co-varying masses and mixing lengths superimposed on the region in the HR diagram that is consistent with observations of HD 140283. We extend this analysis by fitting stellar tracks generated with DSEP to their parameter space and investigating the scope of consistent $\alpha_{\text{MLT}}$ values under variations in DSEP's input physics, based on physical uncertainties in the prescriptions for diffusion and atmospheric boundary conditions.

In table \ref{phx}, we present the sets of mass--$\alpha_{\text{MLT}}$ combinations for which DSEP reproduces the observable features of HD 140283, under four distinct physical configurations. The first sub-table lists the values of $\alpha_{\text{MLT}}$ for a given mass that place the model star's effective temperature and luminosity within the error bounds in C15, as well as the age of the model star at { the track's midpoint of intersection (in luminosity) }. The minimum and maximum temperature and luminosity values along the tracks in the regions of intersection are also given. The second sub-table shows these data for a set of DSEP { tracks implementing a grey model atmosphere}. The last two sub-tables show the same for physical configurations invoking a PHOENIX (standard) model atmosphere, but modified diffusive efficiency: $\eta_{\text{D}}=0.5$ and  $\eta_{\text{D}}=1.5$, respectively. Taking into account the current best constraint on the age of the universe (13.8 Gyr, \citet{Planck}), but relaxing this somewhat to account for uncertainties in stellar dating, only parameter sets that result in intersection ages below $15$ Gyr are shown.

We explore the mass range allowed by DSEP when fitting HD140283's $T_{\text{eff}}$ and $L$. 
We find that in no case is a DSEP track with mass above 0.79 M$_{\odot}$ found to agree with the observational constraints. Mixing lengths ranging from $0.25\alpha_{\odot}$ to $ 1.05 \alpha_{\odot}$ ($\alpha_{\text{MLT}}=0.5$ to $\alpha_{\text{MLT}}=2.0$, with variable grid spacing) are examined.

\begin{figure} % Figure Creevey
\centering
\includegraphics[width=\linewidth]{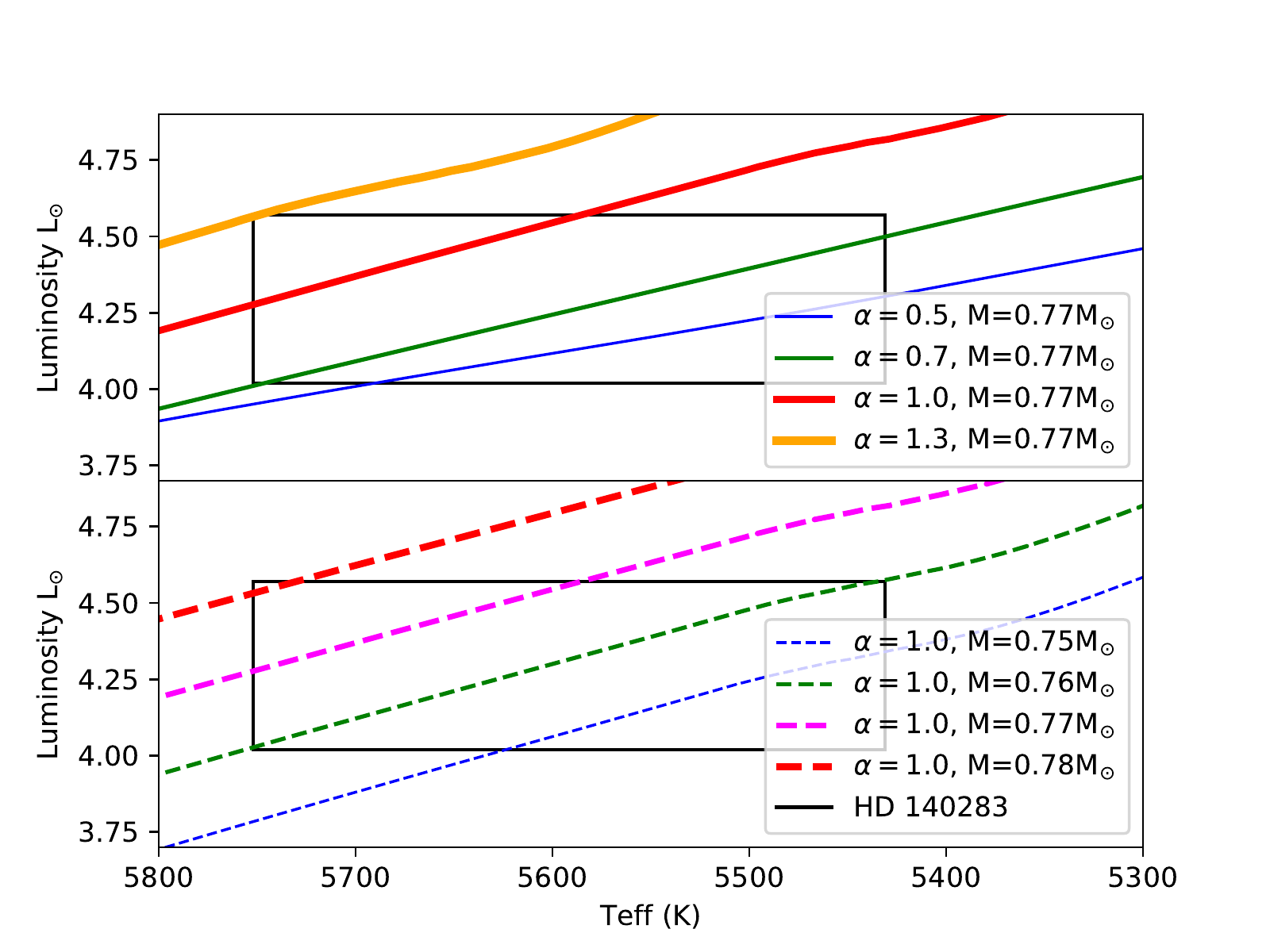}
\caption{
Masses and mixing lengths are fit to HD 140283's temperature and luminosity constraints, provided by C15, using grids of DSEP tracks. The top panel shows the variation in $T_{\text{eff}}$ and $L$ as a function of mixing length for model with mass $0.77 M_{\odot}$. The bottom panel { 
illustrates the effect of mass for fixed $\alpha$,} as listed. All models in both panels use a PHOENIX model atmosphere and $\eta_{\text{D} }=1.0$.
 }
\label{crev}
\end{figure}

Figure \ref{crev} mimics C15's Figure 7, showing analogous mass-$\alpha_{\text{MLT}}$ curves. 
The set of DSEP tracks that intersect with these observational constraints differs somewhat from C15's; in particular, the masses are lower. 
The difference in fitted masses is indicative of differences between CESAM and DSEP, which employ different nuclear reaction rates and atmospheric boundary conditions. Among mixing lengths, { we do find DSEP tracks in agreement with the CESAM tracks} for $\alpha_{\text{MLT}} >0.5\alpha_{\odot}$ ($\alpha_{\text{MLT}}>1.0$), up to and including  $0.89\alpha_{\odot}$ ($\alpha_{\text{MLT}}=1.7$). This is still notably lower than DSEP's { solar-calibrated} mixing length, but this difference falls just short of { the factor of two disparity between $\alpha_{\odot,\text{CESAM}}$ and the fitted value found by C15.} For M$=0.79 M_{\odot}$, however, { agreement is found} using mixing lengths as low as $\sim25$\% of DSEP's solar value.

DSEP produces intersecting curves at every mass between 0.75 and 0.78 M$_{\odot}$ for mixing lengths at C15's fit of $\alpha_{\text{MLT}}=1.0$, or within 10\% of this value. As in their results, DSEP too rules out $\alpha_{\text{MLT}}=2.0$ as a possibility for this star. More to the point, DSEP also cannot produce a consistent curve using its own $\alpha_{\odot}$.

\subsection{Uncertainty in Metallicity}
The metallicity and $\alpha$ element enhancement for HD 140283 are  $\FeH= -2.39\pm0.14$ and $\alphaFe =0.40$, respectively \citep{Creevey}, which imply a best-fitting surface abundance of $Z/X=0.000186^{+0.000108}_{-0.000014}$. Given that metallicity is one of the free parameters in the characterization of HD 140283, it is instructive to examine the impact of adjusting $Z_{0}$. 

\begin{figure} % uncertainty in metallicity; Figure 2
\centering
\includegraphics[width=\linewidth]{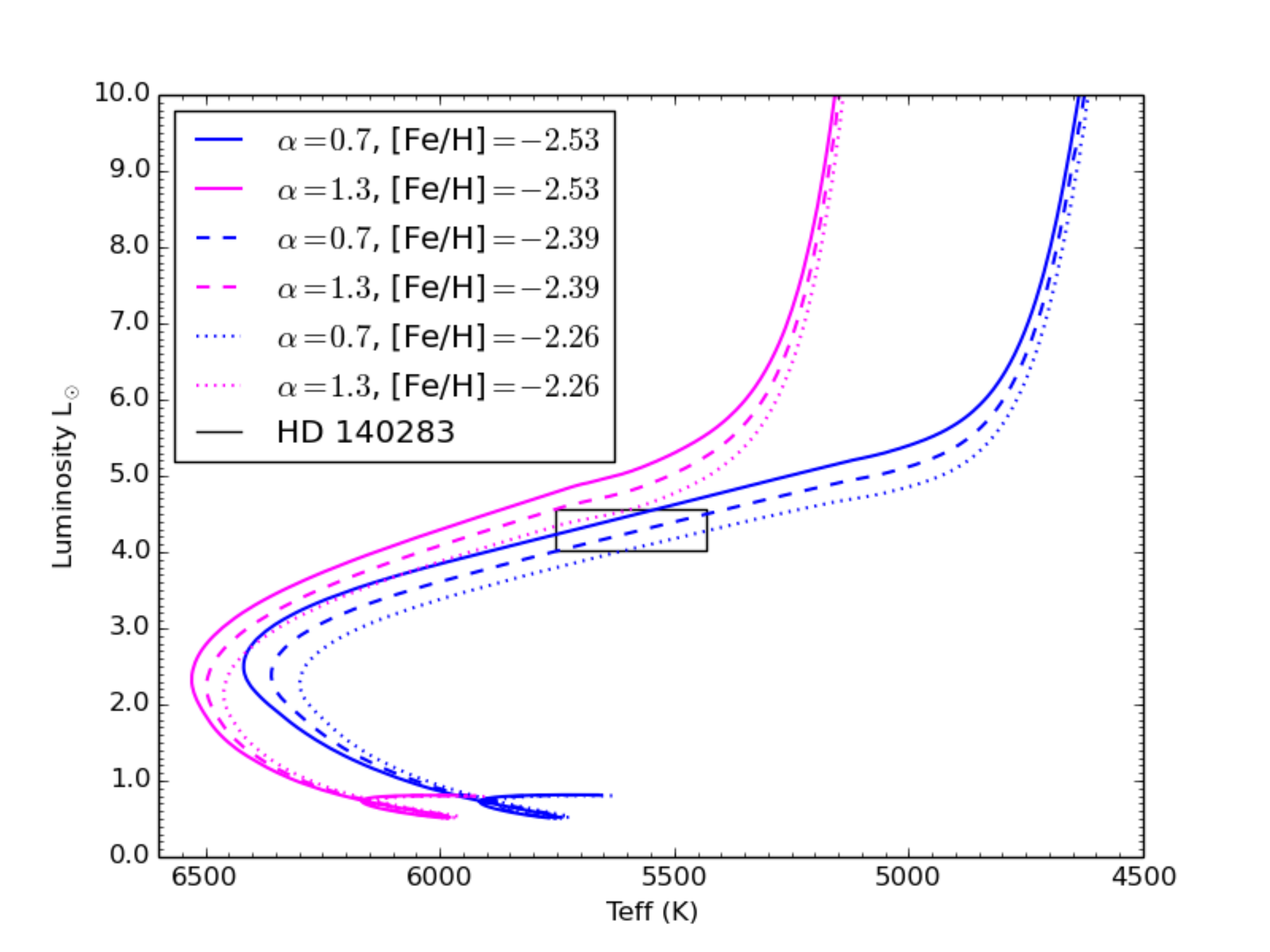} % 7/19/17
\caption{ For two mixing length values $\alpha=0.7, 1.3$, stellar tracks generated with three different input metallicities are shown. The constraints for HD 140283 are shown in black. Tracks invoke a PHOENIX model atmosphere and $\eta_{\text{D}}=1.0$.
}
\label{feh}
\end{figure}

Figure \ref{feh} shows two sets of stellar tracks, grouped by mixing length. For each value of $\alpha_{\text{MLT}}$, tracks are generated for each of $\FeH\in \{-2.53,-2.39,-2.26 \}$ dex (or $Z_{0} \in \{1.057e^{-4}, 1.46e^{-4}, 1.96e^{-4}\}$ for $\alpha$-enhanced models), corresponding to C15's reported value of $\FeH=-2.39$ and taking the uncertainty at its most severe in both directions. Tracks of the same color implement the same mixing length, and tracks of the same line style implement the same $\FeH$. The allowed parameter space for HD 140283 is outlined in black.

In the subgiant region, the effect of varying input metallicity is more dramatic than anywhere else in the HR diagram; raising $\FeH$ by 0.25 dex has an effect on par with that of lowering the mixing length by $\sim0.6$ scale heights. The physical effect of e.g.\ raising $Z_{0}$ is to lower both the estimated effective temperature and luminosity of HD 140283, corresponding to a rise in its fitted age. 
The median range of allowed mixing lengths for fitting HD 140283 at { a zero-age metallicity} of [Fe/H]$=-2.26$ is $\alpha_{\text{MLT}}\sim0.5$-$0.8$. When [Fe/H]$=-2.53,$ the median fit range shifts to $\alpha_{\text{MLT}}\sim 0.9$-$1.7$.

This degeneracy between $\alpha_{\text{MLT}}$ and metallicity should be taken into consideration when constructing the best possible set of parameters. 
We reiterate that we constrain {zero-age} metallicity from observations, which have associated uncertainties, thus making input metallicity itself an adjustable parameter within those uncertainties. Hence, the value derived for metallicity that provides the best fit to HD 140283 is dependent on the values of the other adjustable parameters in the models (like the mixing length), and as such means that possible shortcomings in the model physics may also be absorbed into the {zero-age metallicity.}

{ Because abundances affect the modeling of diffusion, the degree of $\alpha$ element enhancement also affects the chemical evolution of the star. However, though there is observational uncertainty in $\alphaFe$ on the order of 0.15 dex, the impact of varying $\alphaFe$ within this uncertainty is negligible compared to variations in $\FeH$. At, for example, $\FeH=-2.39$, the change in $Z_{0}$ for $\alphaFe=+0.25$ versus $\alphaFe=+0.55$ is less than $0.09\times10^{-4}$. We note that an increase in the degree of $\alpha$-enhancement does raise $Z_{0}$, but constitutes only a small relative increase in metallicity.}

Taking into account this uncertainty, the range of mixing lengths which provide acceptable fits to HD 140283 { is} $\alpha_{\text{MLT}}=0.8$ to $1.3$ across masses, atmospheric boundary conditions, and diffusive efficiency, reaching extreme values of $0.5$ ($\alpha_{\text{MLT}}/\alpha_{\odot}=0.26$, when $\eta_{\text{D}}$ is minimal and mass is maximal) and $1.7$ ($\alpha_{\text{MLT}}/\alpha_{\odot}=0.88$, when mass is minimal and $\eta_{\text{D}}$ is high). 
If a strict age constraint of 13.8 Gyr is applied, the { mid-point of the range of} acceptable values of $\alpha_{\text{MLT}}$ drops to $0.7$. The centermost set of values (with looser age allowance) {corresponds to 42\% to 68\%} of DSEP's $\alpha_{\odot}$ (per physical prescription; see Table \ref{solar}). This is in good agreement with C15's need to invoke a mixing length at roughly half the solar-calibrated value.

%-------------------------------------------------------------------------------------------------
%-------------------------------------------------------------------------------------------------

%-------------------------------------------------------------------------------------------------
%-------------------------------------------------------------------------------------------------

\begin{deluxetable}{ccccc} 
\tablecaption{Matches to HD 140283} 
\tablehead{\colhead{Mass (M$_{\odot}$)}& \colhead{$\alpha_{\text{MLT}}$} &  \colhead{Age (Gyr)} & 
	\colhead{T Range (K)}   & \colhead{L/L$_{\odot}$Range}}
\startdata
	\cutinhead{PHOENIX model Atmosphere,  $\eta_{\text{D}}=1.0$}
0.75  &  0.8  &  14.9  &  5461, 5433  &  4.0, 4.1 \\
  &  0.9  &  14.9  &  5527, 5433  &  4.0, 4.2 \\
  &  1.0  &  14.8  &  5588, 5434  &  4.0, 4.3 \\
  &  1.3  &  14.8  &  5746, 5470  &  4.0, 4.6 \\
  &  1.7  &  14.7  &  5749, 5673  &  4.3, 4.6 \\
\hline
0.76  &  0.7  &  14.3  &  5537, 5432  &  4.0, 4.2 \\
  &  0.8  &  14.3  &  5603, 5432  &  4.0, 4.3 \\
  &  0.9  &  14.2  &  5659, 5434  &  4.0, 4.4 \\
  &  1.0  &  14.2  &  5718, 5434  &  4.0, 4.5 \\
  &  1.3  &  14.1  &  5748, 5558  &  4.3, 4.6 \\
  &  1.7  &  14.1  &  5748, 5748  &  4.6, 4.6 \\
\hline
0.77  &  0.5  &  13.9  &  5627, 5432  &  4.0, 4.2 \\
  &  0.7  &  13.7  &  5701, 5435  &  4.0, 4.4 \\
  &  0.8  &  13.7  &  5747, 5435  &  4.0, 4.5 \\
  &  0.9  &  13.6  &  5748, 5483  &  4.1, 4.6 \\
  &  1.0  &  13.6  &  5749, 5547  &  4.2, 4.6 \\
  &  1.3  &  13.5  &  5747, 5711  &  4.5, 4.6 \\
\hline
0.78  &  0.5  &  13.3  &  5748, 5436  &  4.1, 4.5 \\
  &  0.7  &  13.2  &  5749, 5510  &  4.2, 4.6 \\
  &  0.8  &  13.1  &  5746, 5571  &  4.3, 4.6 \\
  &  0.9  &  13.0  &  5749, 5637  &  4.4, 4.6 \\
  &  1.0  &  13.0  &  5751, 5692  &  4.5, 4.6 \\
\hline
0.79  &  0.5  &  12.7  &  5744, 5574  &  4.4, 4.6 \\
  &  0.7  &  12.6  &  5748, 5689  &  4.5, 4.6 \\
  &  0.8  &  12.5  &  5748, 5736  &  4.5, 4.6 \\\hline
\cutinhead{Eddington grey,  $\eta_{\text{D}}=1.0$}
0.75  &  0.8  &  14.9  &  5477, 5436  &  4.0, 4.1 \\
  &  0.9  &  14.9  &  5538, 5433  &  4.0, 4.2 \\
  &  1.0  &  14.8  &  5598, 5435  &  4.0, 4.3 \\
  &  1.3  &  14.8  &  5747, 5496  &  4.0, 4.6 \\
  &  1.7  &  14.7  &  5748, 5674  &  4.3, 4.6 \\
\hline
0.76  &  0.8  &  14.3  &  5605, 5434  &  4.0, 4.3 \\
  &  0.9  &  14.2  &  5665, 5431  &  4.0, 4.5 \\
  &  1.0  &  14.2  &  5723, 5435  &  4.0, 4.5 \\
  &  1.3  &  14.1  &  5751, 5578  &  4.3, 4.6 \\
  &  1.7  &  14.1  &  5751, 5746  &  4.5, 4.6 \\
\hline
0.77  &  0.5  &  13.9  &  5628, 5433  &  4.0, 4.2 \\
  &  0.7  &  13.7  &  5702, 5433  &  4.0, 4.5 \\
  &  0.8  &  13.7  &  5751, 5435  &  4.0, 4.6 \\
  &  0.9  &  13.6  &  5751, 5503  &  4.1, 4.6 \\
  &  1.0  &  13.6  &  5751, 5562  &  4.2, 4.6 \\
  &  1.3  &  13.5  &  5750, 5717  &  4.5, 4.6 \\
\hline
0.78  &  0.5  &  13.3  &  5745, 5436  &  4.1, 4.5 \\
  &  0.7  &  13.1  &  5748, 5518  &  4.2, 4.6 \\
  &  0.8  &  13.1  &  5751, 5581  &  4.3, 4.6 \\
  &  0.9  &  13.0  &  5747, 5642  &  4.4, 4.6 \\
  &  1.0  &  13.0  &  5751, 5696  &  4.5, 4.6 \\
\hline
0.79  &  0.5  &  12.7  &  5746, 5580  &  4.4, 4.6 \\
  &  0.7  &  12.6  &  5748, 5685  &  4.5, 4.6 \\
  &  0.8  &  12.5  &  5747, 5735  &  4.5, 4.6\\\hline
\cutinhead{PHOENIX model Atmosphere,  $\eta_{\text{D}}=0.5$}
0.76  &  0.8  &  14.5  &  5633, 5436  &  4.0, 4.3 \\
  &  0.9  &  14.4  &  5703, 5432  &  4.0, 4.5 \\
  &  1.0  &  14.4  &  5747, 5439  &  4.1, 4.6 \\
  &  1.3  &  14.4  &  5748, 5613  &  4.4, 4.6 \\
\hline
0.77  &  0.5  &  13.9  &  5627, 5432  &  4.0, 4.2 \\
  &  0.7  &  13.8  &  5720, 5433  &  4.0, 4.5 \\
  &  0.8  &  13.8  &  5747, 5439  &  4.1, 4.6 \\
  &  0.9  &  13.8  &  5751, 5516  &  4.2, 4.6 \\
  &  1.0  &  13.8  &  5748, 5582  &  4.3, 4.6 \\
\hline
0.78  &  0.5  &  13.3  &  5748, 5436  &  4.1, 4.5 \\
  &  0.7  &  13.2  &  5750, 5522  &  4.2, 4.6 \\
  &  0.8  &  13.2  &  5748, 5595  &  4.3, 4.6 \\
  &  0.9  &  13.2  &  5748, 5663  &  4.4, 4.6 \\
  &  1.0  &  13.2  &  5746, 5722  &  4.5, 4.6 \\
\hline
0.79  &  0.5  &  12.7  &  5744, 5574  &  4.4, 4.6 \\
  &  0.7  &  12.7  &  5745, 5692  &  4.5, 4.6 \\\hline
\cutinhead{PHOENIX model Atmosphere,   $\eta_{\text{D}}=1.5$}
0.75  &  0.9  &  14.6  &  5484, 5436  &  4.0, 4.1 \\
  &  1.0  &  14.6  &  5542, 5438  &  4.0, 4.2 \\
  &  1.3  &  14.5  &  5675, 5435  &  4.0, 4.5 \\
  &  1.7  &  14.4  &  5748, 5630  &  4.2, 4.6 \\
\hline
0.76  &  0.8  &  14.1  &  5571, 5431  &  4.0, 4.3 \\
  &  0.9  &  14.0  &  5623, 5436  &  4.0, 4.4 \\
  &  1.0  &  13.9  &  5678, 5431  &  4.0, 4.4 \\
  &  1.3  &  13.8  &  5750, 5508  &  4.2, 4.6 \\
  &  1.7  &  13.8  &  5747, 5694  &  4.4, 4.6 \\
\hline
0.77  &  0.5  &  13.9  &  5627, 5432  &  4.0, 4.2 \\
  &  0.7  &  13.6  &  5682, 5436  &  4.0, 4.4 \\
  &  0.8  &  13.5  &  5727, 5433  &  4.0, 4.5 \\
  &  0.9  &  13.4  &  5748, 5457  &  4.1, 4.6 \\
  &  1.0  &  13.4  &  5747, 5513  &  4.2, 4.6 \\
  &  1.3  &  13.2  &  5746, 5642  &  4.4, 4.6 \\
\hline
0.78  &  0.5  &  13.3  &  5748, 5436  &  4.1, 4.5 \\
  &  0.7  &  13.1  &  5749, 5495  &  4.2, 4.6 \\
  &  0.8  &  13.0  &  5751, 5551  &  4.2, 4.6 \\
  &  0.9  &  12.9  &  5750, 5611  &  4.3, 4.6 \\
  &  1.0  &  12.8  &  5749, 5662  &  4.4, 4.6 \\
\hline
0.79  &  0.5  &  12.7  &  5744, 5574  &  4.4, 4.6 \\
  &  0.7  &  12.6  &  5745, 5680  &  4.5, 4.6 \\
  &  0.8  &  12.5  &  5748, 5723  &  4.5, 4.6 \\
 \hline \\[4pt]
\enddata
\tablecomments{Matches to \citet{Creevey}'s parameter space for HD 140283, using a grid of DSEP stellar tracks spanning a mixing length range $\alpha_{\text{MLT}}=0.5\text{--}2.0$ and a mass range $0.74\text{--}0.79$ M$_{\odot}$, with varying physical prescriptions as indicated.}
\label{diff05}
\label{ttau}
\label{phx}
\label{diff15}
\end{deluxetable}

%-----------------------------------------------------------------------------------------

\section{Degeneracies in Age, Mass, and Mixing Length}
As others (e.g.\ \citet{Tayar}) have demonstrated, the mixing length can heavily impact the predicted age of a model star. For a stellar track designed to reproduce a specific mass, effective temperature, and luminosity (as with HD 140283), a lower mixing length raises the age at which a model of given mass will reproduce the star's temperature and luminosity. 

\begin{figure} % Figure age_mass_ML:   redo with fewer lines
\centering
\includegraphics[width=\linewidth]{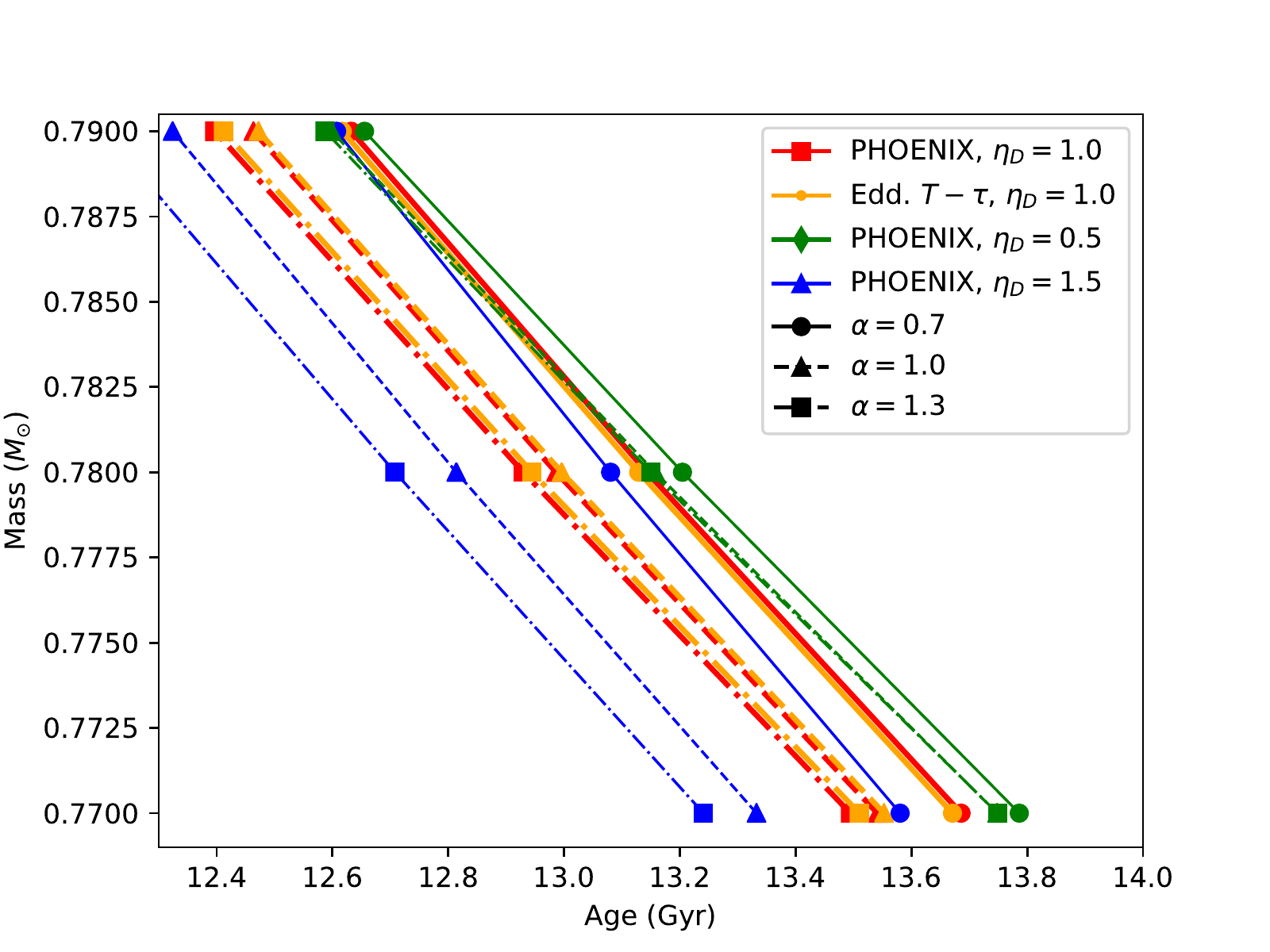} % 8/11/17
\caption{ 
Input track mass (M$_{\odot}$) is shown as a function of intersection age (Gyr) for mixing lengths $\alpha_{\text{MLT}} \in \{ 0.7,1.0,1.3\}$. The value of $\alpha_{\text{MLT}}$ within a set of tracks (grouped by color) decreases from left to right; e.g.\ smaller mixing lengths intersect at higher ages. The red curves with square markers correspond to models generated with the standard prescription (Phx atmosphere, $\eta_{\text{D}}=1.0$). The yellow curves with circular markers {correspond to models} generated with a grey model atmosphere and $\eta_{\text{D} }=1.0$. Curves marked with green diamonds and blue triangles correspond to curves generated with PHOENIX model atmospheres and $\eta_{\text{D}}=0.5$, 1.5, respectively.  }
\label{AML}
\end{figure}

Figure \ref{AML} demonstrates the effect of mixing length { on the age of present-day models of HD 140283 as a function of mass.
The figure shows a few mixing length values for each of the four physical configurations as a function of age, over a subset of the range of masses for which DSEP tracks agree with C15's observational constraints: 0.74 to 0.79 M$_{\odot}$. Only mass and $\alpha_{\text{MLT}}$ combinations which produce intersection between the stellar track and C15's observational constraints on temperature and luminosity are shown.}

Tracks generated under the same physical configuration are shown in the same color and line style, in order to emphasize the impact of the input physics---primarily $\eta_{\text{D}}$---on tracks irrespective of particular mixing length and mass values. Mixing lengths within each of these sets of tracks increases from right to left {(e.g.\ models with lower mixing lengths, for a given mass, are older for a fixed point in $T_{\text{eff}}$-$L$ space), demonstrating that models with the same mass but lower mixing lengths evolve more slowly.}

As has been demonstrated in the literature (e.g.\ \citet{Chab95}), the impact of altering diffusion is much more evident than that of changing the model atmosphere. Taking the red set of curves to represent the canonical, or ``default'' configuration (PHOENIX model atmosphere, $\eta_{\text{D}}=1.0$),  we see that when diffusion is made half as efficient (green curves),  model stars of a given mass are older. We also find that the mixing length has less of an impact on the star's age overall in the lower-diffusion {model}; this is { inferred from the tighter spacing between} the mixing length curves using suppressed diffusion relative to the sets of models with standard or enhanced diffusion. 

As discussed in Section 2, this can be understood in terms of the conjugate effects of mixing in convective regions and efficiency of heavy element diffusion. Because lessening the efficiency of diffusion mimics lowering $\alpha_{\text{MLT}}$, one should anticipate that lowering $\eta_{\text{D}}$ lessens  $\alpha_{\text{MLT}}$'s impact overall. That the same mixing lengths create older stars under low-diffusion conditions speaks to the same degeneracy: with a lower degree of chemical mixing in the stellar interior comes longer burning cycles, and hence higher ages.

 Using a grey model atmosphere over the PHOENIX model atmosphere does not affect the median age of intersection appreciably or consistently;  the difference between a track using the canonical configuration and the Eddington track is only noticeable on the order of 0.05 Gyr. In contrast, there is nearly an 0.7 Gyr difference between a canonical track and a track using $\eta_{\text{D}}=0.5$, for the models of lowest mass.

{It is important also to note the large effect on the ages of the stellar models due to changes in the mixing length. Models of a given mass can differ in age by up to half a billion years through a change of
$\alpha_{\text{MLT}}=\alpha_{\odot}$ to $\alpha_{\text{MLT}}=0.5 \alpha_{\odot}$. We note also that } the impact of the mixing length on the age of the model is not linear with $\alpha_{\text{MLT}}$; it is proportionally greater for lower-mass stars. 
This trend is due to the fact that the convective envelope takes up a proportionally larger region of the stellar interior for low-mass stars.

Figure \ref{AML} explicitly quantifies the impact of changes in $\alpha_{\text{MLT}}$ relative to { changes in $\eta_{\rm D}$ and $T(\tau)$.} When diffusive efficiency is maximized, a change of 0.3 scale heights in $\alpha_{\text{MLT}}$ results in an age difference of nearly { 0.3 Gyr}, for models at the high end of the mass spectrum and mixing lengths in the lower regime. At the opposite end of both spectra, the impact of mixing length on age is virtually non-existent for low $\eta_{\text{D}}$. Irrespective of differences in physical inputs, $\alpha_{\text{MLT}}$'s impact on age likewise diminishes for lower masses and lower values of $\alpha_{\text{MLT}}$. In addition to these coupled effects, we also reiterate that the impact of mixing length on age is degenerate with metallicity { near the TO and along the subgiant branch} (see Figure \ref{feh}).

{ Degeneracies among parameters are a notorious source of uncertainty in stellar age determinations from model fitting;}
the source of frustration being that we { have, until recently, lacked the ability to place robust observational constraints on a star's interior properties (asteroseismology has been successful in this regard, but not for a broad spectrum of stars; some alternative HR-dependent methods for age determinations have also been investigated---see e.g.\ \citet{Nataf}, \citet{me} and other studies focusing on the red giant branch bump).}	
For now, we quantify these degeneracies as best we can, and maintain strict awareness of the pitfalls of working with multiple free parameters.

%-------------------------------------------------------------------------------------------------
%-------------------------------------------------------------------------------------------------

\section{Fitting Other Regimes}
Although HD 140283 provides the only empirical test we have, it is arguably more reasonable to generalize our findings about its best-fitting mixing length to all highly metal-poor subgiants than it is to generalize the solar mixing length to all stars. That is to say, it is not ideal. One way we can expand our knowledge, however, is by { studying mixing length fits to stars of similar composition as HD\,140283, but for a range of masses and ages.}

\subsection{Best Fits to M92}
While our findings agree with the need to implement a mixing length significantly smaller than the solar-calibrated value to fit subgiant HD 140283, { we also ask whether the mixing length should be adaptive depending on atmospheric parameters, and hence on age of the star, as supported by $\alpha_{\rm MLT}$-calibrations against 3D simulations \citep{Freytag99, Trampedach, MagicMLT}.} For this reason, we examine the fits of stellar isochrones over a range of mixing lengths to observations of stars in different evolutionary phases, beginning with the red giant branch. 

Though there is no other very metal-poor star with observational characterization as extensive as HD 140283's, 
{ a globular cluster with very similar metallicity allows for the examination of $\alpha_{\rm MLT}$ for subgiants, red giants, and for a significant mass range along the main sequence.}

The data for M92 are provided by \citet{Sara}, who present a reddening and distance modulus of $E(B-V)=0.02$ and $m-M_V=14.7$. When fitting M92, we calibrate our own reddening and distance modulus values in accordance with DSEP's isochrones. The values we find vary slightly depending on the physical inputs used in DSEP (see discussion in Section 6.1.1 for more details).

To determine the best-fitting mixing length, we create isochrones comprising grids of stellar tracks generated for mixing lengths spanning $\alpha_{\text{MLT}}=0.5$ to $\alpha_{\text{MLT}}=3.0$, or roughly 0.25 to 1.5 times the solar mixing length, with variable grid spacing (higher  $\alpha_{\text{MLT}}$ values require denser sampling to account for the reduced $T_{\text{eff}}$ sensitivity at higher mixing lengths; {see \ref{twoML} }). 
{ We once again consider isochrones between the ages of 8--16 Gyr, in increments of 1 Gyr (and likewise find fits only between the ages of 11 to 15 Gyr). } For each mixing length, we then visually determine the isochrone and age that best fit M92. As with HD 140283, we determine the best-fitting mixing length under four physical prescriptions encompassing two model atmospheres and three values of $\eta_{\text{D}}$. 

%M92PHX
Table \ref{M92only} lists the mixing lengths, normalized mixing lengths (the mixing length divided by the solar-calibrated mixing length corresponding to the relevant physical prescription), and ages of each best-fitting isochrone. Figure \ref{m92fit} shows an { example of the fits of isochrones spanning the aforementioned age range to M92 for models generated} using a grey model atmosphere and a mixing length of $\alpha_{\text{MLT}}=1.65$ (the color calibration scheme of \citet{VC}---hereafter ``VC''---is used to transform all isochrones from $T_{\text{eff}, L}$ to colors and magnitudes unless otherwise specified; Section \ref{secsith} gives more detail).

The data reflect the same trend first explained in the solar mixing length calibration: the best-fitting $\alpha_{\text{MLT}}$ values { are higher for higher values of $\eta_{\text{D}}$.} { In fits to M92, the grey atmosphere models require slightly lower values of $\alpha_{\rm MLT}$ compared to models using Phx atmospheres, but this is not the case in the fits to HD 140283 or to any of the metal-poor parallax stars (see Section 6.2). }
The best-fitting mixing length values are not different enough among the first three physical prescriptions to merit extensive physical speculation. What is much more important to recognize is that $\alpha_{\text{MLT}} < \alpha_{\odot}$ by a meaningful degree ($\sim10$\%) for every prescription.

In Figure \ref{m92fit}, we see that the main sequence turn-off (TO) and subgiant branch are the most parameter-sensitive regions of the fit to M92; the best-fitting age is determined exclusively by the TO. { The tracks along the main sequence and the red giant branch, on the other hand, show virtually no effect of even a 5\,Gyr age difference} for a given $\alpha_{\text{MLT}}$. However, the red giant branch does display extreme sensitivity to $\alpha_{\text{MLT}}$ { itself---Figure \ref{twoML} shows a range of 13 Gyr isochrones, }
spanning $\alpha_{\text{MLT}}=0.7$ to $\alpha_{\text{MLT}}=1.95$ (roughly $\alpha_{\odot}$), with otherwise identical parameters: a PHOENIX model atmosphere and  $\eta_{\text{D}}=1.0$. In summary, the length of the subgiant branch and the curvature of the red giant branch are the decisive features in determining both the best-fitting mixing length and the best-fitting age.

We find that, relative to the acceptable ranges for HD 140283, higher values of $\alpha_{\text{MLT}}$ provide the best fits to M92. For all physical prescriptions, the best-fitting mixing length hovers at roughly 90\% of its respective solar-calibrated value. While this is not a { large} deviation from $\alpha_{\odot}$, the constancy of the normalized mixing lengths { across widely different physics---both in terms of the range of evolutionary stages captured by fits to a globular cluster and in terms of variations in diffusion and atmospheric boundary conditions---is worth recognizing.}
The consistency is noteworthy especially given that the isochrones are fitting multiple evolutionary phases simultaneously by fitting a large cluster of stars, though the subgiant and red giant branches display the most parameter sensitivity and functionally dictate the fit. Paying particular attention to the red giant branch, we conclude that a slightly subsolar value of $\alpha_{\text{MLT}}$ is required to reproduce the observed properties M92 as accurately as possible, and suggest that lower values may be required to fit metal-poor red giants in general.

Our findings for M92 differ slightly from results reported by \citet{Tayar}, who would predict a mixing length of $\sim 1.6$, { or 0.93 $\alpha_{\odot,\text{YREC}}$ (using the YREC stellar evolution models),} for M92 based on their $\alpha_{\text{MLT}}$-[Fe/H] trend. We note, however, that their results are based on stars with much higher metallicities than what we consider in this paper: [Fe/H]$\sim -0.5$ dex in \citet{Tayar} versus [Fe/H]$=\sim -2.3$ dex here.

\subsubsection{Additional Fitting Parameters}
In the case of M92, the determination of the best-fitting isochrone is also contingent on the distance modulus and reddening adopted for the observational data. { We find that applying the values for distance modulus and reddening reported in \citet{Sara}, $m-M_v=14.7$ and $E(B-V)=0.05$, respectively, to the M92 data provide the best fit to a model using the default physical configuration and $\alpha_{\text{MLT}}=1.8$.
We use \citet{Sara}'s values as an initial baseline throughout our M92 analysis, but manually adjust both parameters as needed to improve the fits case by case. In Figure \ref{m92fit}, for example, the optimal distance modulus varies from $\sim14.5$ to $14.8$, and the optimal reddening varies from $0.047$ to $0.065$, with younger isochrones requiring the highest values in both parameters. 
In Figure \ref{twoML}, the optimal distance modulus is found to be $m-M_v=14.7$ in all cases, and the optimal reddening ranges from $0$ to $0.6$.

Figures \ref{m92fit} through \ref{parfit} are shown in terms of apparent rather than absolute magnitude so that we can demonstrate the adjustments in $m - M_V$ and $E(B-V)$ to each isochrone individually.  
In the cases where $\alpha_{\text{MLT}}$ is smallest in Figure \ref{twoML}, reddening values below zero would provide better agreement between the isochrone and the M92 data; however, we restrict to $E(B-V)>0$ for the moment.
 We {compare} otherwise physically identical isochrones of different ages as well,}
allowing for some flexibility in the upper age limit (15 Gyr) to account for the age-$\alpha_{\text{MLT}}$ covariance.

{ Figure \ref{multiML} shows the fits of isochrones with a variety of diffusive efficiencies and atmospheric boundary prescriptions, and again in each case, the age of the isochrone, distance modulus, and reddening are varied to optimize the fit. $\alpha_{\text{MLT}}=1.8$ remains fixed for all isochrones in this figure. Table \ref{M92table} provides more detail on these parameters. In this case, we do allow for negative reddening values in our optimization for the purpose of demonstration. 
{We acknowledge that, for example, the negative reddening values found to optimize the fit at $\alpha_{\text{MLT}}=1.1$ and $\alpha_{\text{MLT}}=0.9$ are not physically valid. 
This indicates that these mixing lengths are unsuitable for the prescriptions specified; otherwise reasonable mixing length values can, of course, be ruled out by physical violations besides a poor fit in magnitude-color space.} Similarly, it follows that reasonable $\alpha_{\text{MLT}}$ values should not be ruled out by poor color transformations alone.

%------------------- ADDING DM AND REDDNING-----------------------
\begin{table*} 
\centering 
\caption{Mixing Length Values of Best-Fitting Isochrones to M92} 
\begin{tabular}{ l l l l  l l l l  l}  
\hline\hline
%% updated 7/24/17
 { Object  }& {Model Atm} & {$\eta_{\text{D}}$} & $\FeH$ & 
 $m - M_V$ & $E(B-V)$ &
 {$\alpha_{\text{MLT}}$}  & {$\alpha_{\text{MLT}}/\alpha_{\odot}$} & {Age }  \\ \hline
M92			&Phx 			& 1.0		&-2.4	& 14.6	&	0.058	& 1.75	&0.9087		&13 \\		
			& grey			& 1.0		&-2.4	& 14.7  &   0.058	& 1.65 	&0.906		&12 \\
			&Phx 			& 0.5 	 	&-2.4 	& 14.7 & 	0.058 & 1.7 	&0.929		&13 \\
			&Phx 			& 1.5 		&-2.4 	& 14.6 & 0.058	& 1.75 	&0.885		&13  \\
\hline
\end{tabular}
\tablecomments{
We show the mixing length value and other fit parameters for the best-fitting isochrones to the M92 data for each physical prescription.
The column $\alpha_{\text{MLT}}/\alpha_{\odot}$ describes the best-fitting mixing length normalized by the solar-calibrated mixing length value for the associated configuration; e.g.\ values of $\alpha_{\text{MLT}}$ corresponding to models with an Eddington/grey atmosphere and $\eta_{\text{D}}=1.0$ are divided by $\alpha_{\odot}=1.8205$ (see Table \ref{solar}).
We also provide the reddening and distance modulus applied to the M92 data in each case, which we allow to vary slightly from \citet{Sara}'s values of $m -M_V=14.7$ and $E(B-V)=0.05$. We note that the distance modulus and age are degenerate parameters, but the adjustments in distance modulus are minor compared to the age resolution among the isochrones (1 Gyr).
}
\label{M92only}
\end{table*}

%-----------------------------------------------------------------------------------

\begin{table} 
\centering 
\caption{Adjustable Parameters for Fits to M92 (Fig. \ref{multiML})}
\begin{tabular}{ l l  l l l l  }  
\hline\hline
{$\alpha_{\text{MLT}}$}  & { Atmosphere  }& {$\eta_{\text{D}}$} & {$m-M_V$} &  {$E(B-V)$} & {Age (Gyr)}  \\ \hline
%3.0			&Phx 			& 0.0		&-2.25	& 1.75			&13 \\		
3.0			&PHOENIX		& 0.5		& 14.9		& 0.13 		&12 \\
2.5			& Grey 			& 1.0 	 	& 14.8 		& 0.10 		&12 \\
1.8			&PHOENIX 		& 1.0 		& 14.7 		& 0.05 		&13  \\
1.1			&PHOENIX 		& 1.0 		& 14.4 		& -0.05 	&15  \\
0.9			& Grey 			& 1.0 	 	& 14.5 		& -0.07 	&15 \\
\hline
\end{tabular}
\tablecomments{
For each isochrone of specified mixing length, the set of observational and theoretical parameters that gives its best possible fit to M92 is presented. Distance modulus, reddening, and age are permitted to vary in constructing a fit. $m-M_V$ { is the} fitted distance modulus, and $E(B-V)$ gives the fitted reddening. { We reiterate that negative reddenings are unphysical, and they are only presented to demonstrate the optimal fit parameters. }
}
\label{M92table}
\end{table}

%-------------------------------------------------------------------------------------------------
%-------------------------------------------------------------------------------------------------
\begin{figure} % figure: demonstrates grid fit to M92: OK 6/24/17
\centering
\includegraphics[width=\linewidth]{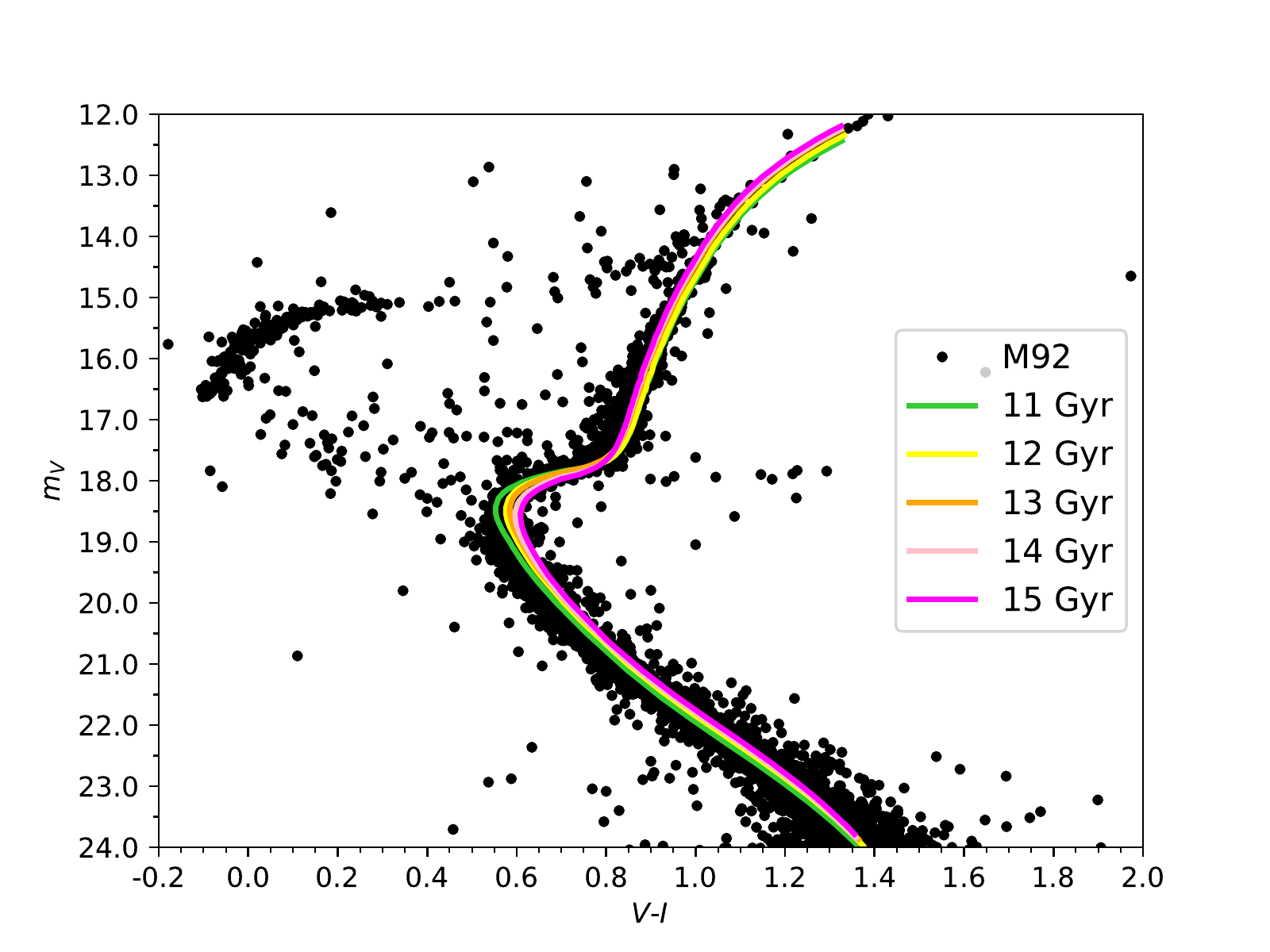}
\caption{
The fits of five isochrones spanning ages 11 to 15 Gyr, in increments of 1 Gyr, are shown against M92 { in terms of apparent magnitude, to allow for variations in distance modulus and reddening to be applied to the isochrones.} All models shown here invoke a grey model atmosphere, $\eta_{D}=1.0$, $\alpha_{\text{MLT}}=1.65$, and VC color calibration. 
}
\label{m92fit}
\end{figure}
%-------------------------------------------------------------------------------------------------

%-------------------------------------------------------------------------------------------------
\begin{figure} % figure: demonstrates grid fit to M92: OK 6/24/17
\centering
\includegraphics[width=\linewidth]{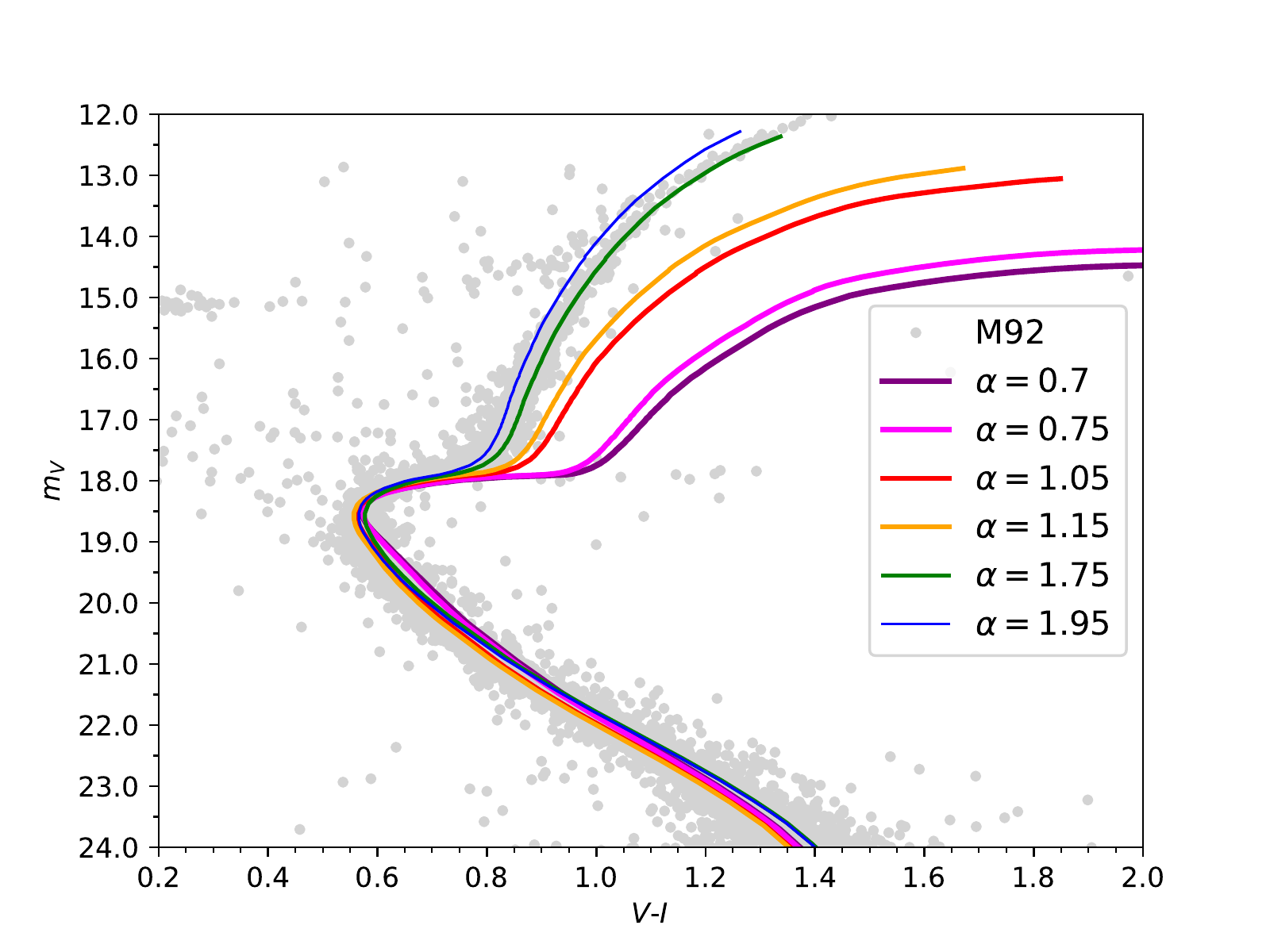}
\caption{
Six isochrones, each of age 13 Gyr, are generated with different mixing lengths and shown against M92 for reference. Each isochrone in the Figure invokes the same physical prescription (PHOENIX model atmosphere, $\eta_{\text{D}}=1$) and color calibration (VC). 
{ The M92 data are once again presented in apparent magnitudes to allow for individual tailoring in distance modulus and reddening, applied to the isochrones. $m-M_V=14.7$ in all cases; $E(B-V)$ ranges from $0$ to $0.06$.}
The mixing length's greatest effect is on the shape of the isochrone, in particular on the length, and hence duration, of the subgiant branch. The goodness of fit is determined primarily by the isochrone's alignment with the red giant branch; highly subsolar values of $\alpha_{\text{MLT}}$ fit this region quite poorly in the case of M92.  
}
\label{twoML}
\end{figure}
%-------------------------------------------------------------------------------------------------

%-------------------------------------------------------------------------------------------------
\begin{figure} % figure: multiple mixing lengths for one config
\centering
\includegraphics[width=\linewidth]{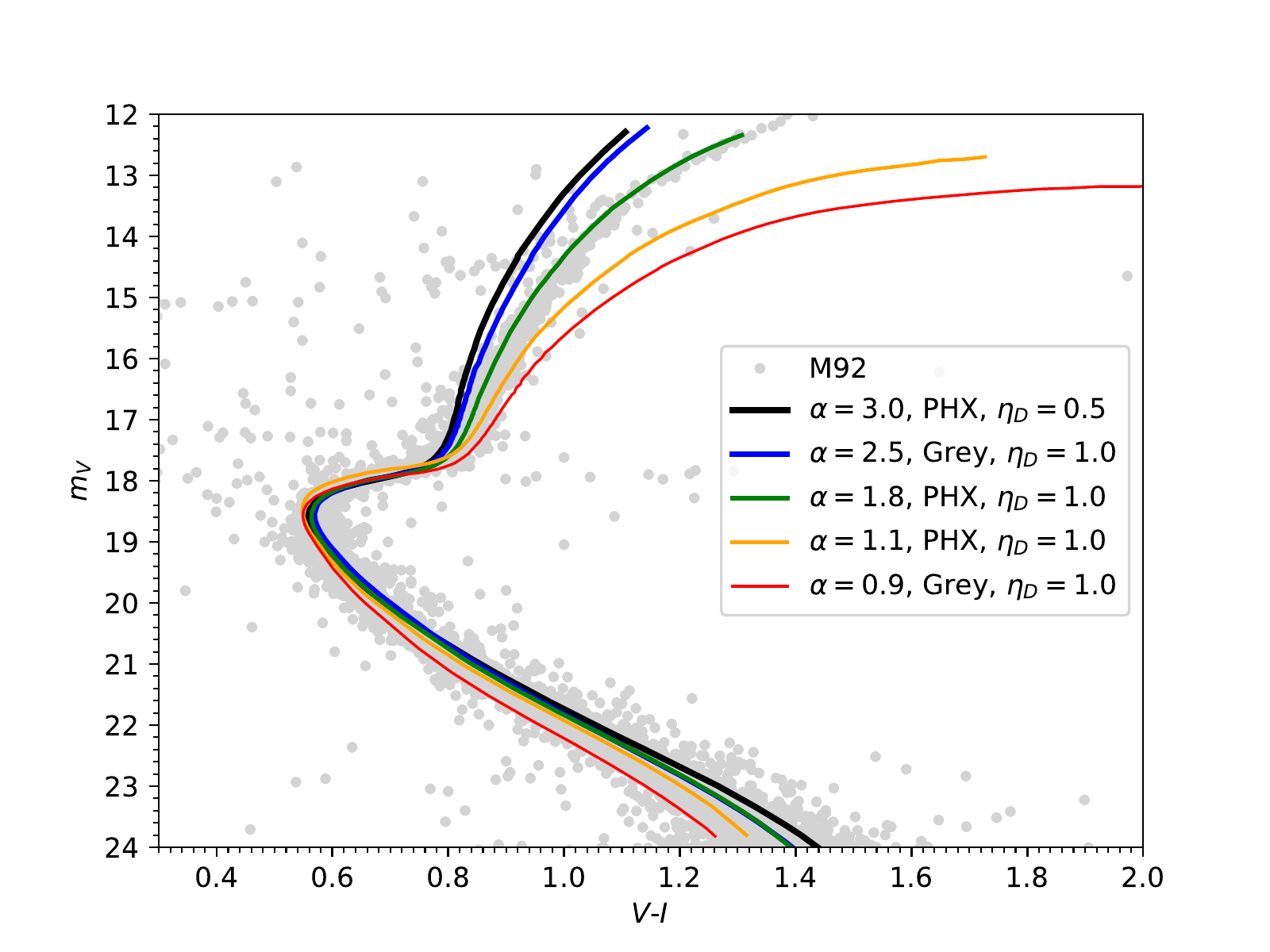}
\caption{ 
A sample of isochrones at their respective optimized fit parameters, given in Table \ref{M92table}, is shown for a selection of configurations and mixing lengths. 
The distance modulus and reddening are also fit individually and given in Table \ref{M92table}. 
A VC color calibration is applied to all. 
}
\label{multiML}
\end{figure}
%-------------------------------------------------------------------------------------------------

\begin{table} 
\centering 
\caption{Properties of Fitted Objects }
\begin{tabular}{ l c c     c c  l }  
\hline\hline
{ Name } & V & $V-I$  &  $\FeH$ 
% &[$\alpha$/Fe]  0.29,0.19,0.23
%& $\alpha_{\text{MLT}}$
 & Reference  \\ \hline
HD 140283  	&  7.21  &  0.0   & $-2.46$  &\citet{Creevey} \\
HIP 46120  	& 10.12  & 0.752  &  $-2.22$  & \citet{Chab2017} \\
HIP 54639  	& 11.38  & 0.914  &  $-2.50$  & \citet{Chab2017} \\
HIP 106924  & 10.36  & 0.803  &  $-2.23$  & \citet{Chab2017} \\
WOLF 1137  	& 12.01  & 0.85   &  $-2.53$   &\citet{Erin}  \\
M92  		&  	-	 &  - 	  &  $-2.24$  &\citet{Sara}\\
\hline
\end{tabular}
\tablecomments{ The observed properties of the objects we consider in this study are presented with the relevant reference. The reddening and distance modulus provided by \citet{Di} for M92 are $E(B-V)=0.025$ and $m-M=14.58$, respectively. These values differ slightly from the values found to best align with DSEP's isochrones; they vary with physical configuration, but average closer to the reddening and distance modulus provided for M92 in \citet{Sara}: $E(B-V)=0.05$ and $m-M=14.7$, respectively)
}
\label{prop}
\end{table}

\subsection{Best Fits to Low-Metallicity Parallax Stars}
Having demonstrated that subsolar values of $\alpha_{\text{MLT}}$ produce better models of metal deficient stars in their post-TO phases, we move on to an investigation of the main sequence. Until this point, we have also only examined { stars} with nearly identical metallicities. We now consider four highly metal-poor, main sequence stars with two degrees of severe metal depletion: $\FeH\sim-2.2$ and $\FeH\sim-2.5$. Taking into account their $\alphaFe$ abundances, the less deficient pair of stars shares essentially the same $Z_{0}$ as HD 140283 and M92, and can be fitted by the same tracks and isochrones. For the more metal-deficient pair, isochrones are generated using tracks with approximately half the $Z_{0}$ value used in fitting HD 140283 and M92.

The four stars we fit are HIP46120, HIP54639, HIP106924, and  WOLF1137. Each is a highly metal-poor, main sequence subdwarf with well-determined parallax, and hence well-constrained absolute magnitude. Their properties are summarized in Table \ref{prop}. Metallicities, reddenings, and photometric observations for these stars are provided by \citet{Erin}. Parallaxes were independently determined from Gaia TGAS \citep{TGAS}.
Further details of the observations are given in \citet{Chab2017} and \citet{Erin}.

For HIP 46120 and HIP 106924, we sample the same grid of isochrones used to fit M92, due to their similar $Z_{0}$ values, taking into account $\alpha$ element enhancement ($Z_{0}\sim1.5 \times 10^{-4}$). For HIP 54639 and WOLF 1137, we create a similar grid using $Z_{0}\sim7.8 \times 10^{-5}$. Grid spacing { in $\alpha_{\text{MLT}}$} is on the order of $\sim0.05$ { scale heights,} which {translates to lower $T_{\text{eff}}$ resolution} in the low-$\alpha_{\text{MLT}}$ regime.

Figure \ref{parfit} shows a sampling of isochrones that fit HIP 46120 and WOLF 1137 for a range of parameter combinations. Cases considered are, once again, PHOENIX versus grey model atmospheres; enhanced, suppressed, and standard diffusive efficiency ($\eta_{\text{D}}=0.5,1.0,1.5$); isochrone ages spanning 11 to 15 Gyr in increments of 1 Gyr; and mixing lengths between 25\% and 150\% of the solar value (per case). 
Table \ref{bestfitVC} gives the best-fitting values of $\alpha_{\text{MLT}}$ under each set of conditions. HD 46120 and HD 106924 are fit only by isochrones with $\FeH=-2.25$, and likewise for HIP 54639 and Wolf 1137 with  $\FeH=-2.5$.

While the vast majority of best-fitting mixing lengths are subsolar across all physical cases, the $\alpha_{\text{MLT}}$ values are highly segregated by star; there is no generalized ``main sequence trend'' that we can infer from these data. HIP 46120 and Wolf 1137 are found to be fit best by normalized mixing lengths between M92's $\alpha_{\text{MLT}}/\alpha_{\odot}\sim 0.9$ and $\alpha_{\odot}$. The normalized mixing lengths are not, however, nearly as consistent across physical configurations as they were found to be for M92. And, in a few cases, $\alpha_{\text{MLT}}$ is found to be slightly supersolar.

The remaining subdwarfs tell a different story. A mixing length just over half the solar value is required to fit HIP 106924, and a mixing length one-third the solar value fits HIP 54639. From the perspective of empirical calibration, this means that the solar mixing length is ``off'' by a factor of 2 (or 3!) for physical prescriptions meant to { describe} these stars---{ quite the lack of accuracy for a parameter which has, historically, been calibrated to 5 digits' precision.}

Although there is high variance among the best-fitting mixing lengths across stars ($\alpha_{\text{MLT}}=0.5$ to 2.1), the range of normalized mixing lengths per star is small; $\alpha_{\text{MLT}}/\alpha_{\odot}$ varies by a maximum of $\sim10$\% across changes in model atmosphere and $\eta_{\text{D}}$. It is worth noting that, if separating the four subdwarfs into groups, one with an average $\alpha_{\text{MLT}}/\alpha_{\odot}>0.7$ and one below $0.7$, a star of each metallicity appears in each group. This counters the possible suspicion that metallicity may be the cause of the divergence.

{This lack of sensitivity of the mixing length along the main sequence can be at least partially understood in terms of knowledge from theoretical stellar evolution. The main effect of changing the mixing length is to change the entropy of the asymptotic adiabat of the deep convection zone. This in turn changes the temperature of the bottom of the convection zone and the location at which the adiabat crosses the radiative stratification of the interior. 
Lower values of $\alpha_{\text{MLT}}$ correspond to higher entropies of the adiabat, meaning the convection zone will start at a deeper point in the interior. However, the fractional size of the convection zone does not change much, which means the star as a whole will be smaller.
Hence, a higher $\alpha_{\text{MLT}}$ gives only a slightly smaller convection zone as fraction of radius, but it gives a significantly smaller radius of the star overall \citep{Dalsgaard97}. 
This change in the location of the bottom of the convection zone accounts for the difference in stellar radius observed with varying the mixing length.}

Figure \ref{twoZ} demonstrates the difference between isochrones that invoke $\FeH=-2.5$ versus $\FeH=\allowbreak -2.25$. As with M92, age fitting is also done visually, but the determination of a best-fitting age is much more difficult for these stars. As Figure \ref{twoZ} demonstrates, the effects of uncertainty in age and metallicity are degenerate. Explicitly, the range of colors (i.e.\ temperatures) covered by an age span of 11--15 Gyr is on par with the range allowed by an uncertainty in metallicity of less than 10\%---a more conservative estimate than what has been adopted as theoretical uncertainty in the DSEP stellar evolution code in the past \citep{Chab06, me}, and a more conservative bound than, for example, the uncertainty C15 places on HD 140283. One should not, in other words, ascribe a great deal of physical significance to the selection of a best-fitting age along the main sequence, in light of more consequential sources of uncertainty.

Though we can make no statement about an all-encompassing, empirical value of $\alpha_{\text{MLT}}$ that unilaterally fits the main sequence, we have demonstrated two important details. First, the best-fitting values of $\alpha_{\text{MLT}}$ found for these stars, across a plethora of parameter fields, are, once again, subsolar. At their least egregious, they hover just below $\alpha_{\odot}$; at their worst, $\alpha_{\text{MLT}} < 0.3 \alpha_{\odot}$. Second, the fact that such large changes in $\alpha_{\text{MLT}}$ are demanded by stars with such similar characteristics 
supports the fact that we would do well to re-evaluate our stellar models and their connection to the observable properties of stellar atmospheres. A powerful approach to this is asteroseismology, which offers us a window into stellar interiors with very different model-dependencies than classical analysis of stellar observations.

These findings provide a strong case for the re-observation of these stars, to verify their colors and temperatures, as well as
motivate the acquisition of more observations of similar stars. 
Although there are no other metal-poor stars with observed radii (besides HD 140283), increasing the sample size of single, metal-poor, main sequence stars with well-determined parallaxes would increase the robustness of our model tests in similar fashion. There are a number of candidate stars with existing high-resolution spectroscopic abundances, known { metal depletion} [Fe/H]$<0.6$, and TGAS parallaxes, and mixing length fits to a large number of such observations would invariably provide more insight. Expanding the data set in this way constitutes the first step toward confirming the existence, or lack, of a mixing length trend along the main sequence.

\subsection{Impact of Color Calibration}
\label{secsith}
The final computational step in fitting isochrones to observational data is the transformation from luminosity and temperature to magnitudes and colors. { This requires that we apply a color calibration to DSEP's (theoretical) output.}

{ We consider two color calibrations in this analysis: the {semi-empirical} method of \citet{VC} (hereafter VC)---assumed as our default calibration and used for all isochrones presented thus far---and the method of \citet{Hauschildt} (hereafter PHX). 
Despite the fact that use of the PHX calibration is most self-consistent with our use of the PHOENIX model atmospheres,
we choose to apply the VC transformation as our default because it is semi-empirical.
{For the cooler stars discussed in this paper, the \citet{VC} color transformations are based upon the model atmospheres of \citet{BellGust78} and \citet{VanBell85}, adjusted to conform to observational constraints (see \citet{VC} for a thorough discussion of how this is implemented).}
We likewise find that the VC tables lead to better fits to observations than do the PHX tables. 
We explore the effect of using PHX's purely theoretical transformation for the sake of completeness.}

{ Tables \ref{M92PHX} and \ref{bestfitPHX} give the best-fitting $\alpha_{\text{MLT}}$ values for isochrones calibrated instead according to synthetic PHX colors, for M92 and the subdwarfs, respectively. Comparing Tables \ref{M92only}, \ref{M92PHX}, \ref{bestfitVC}, and \ref{bestfitPHX}, what stands out immediately is the higher absolute and relative values of $\alpha_{\text{MLT}}$ when a PHX color calibration is used. The best-fitting values under this transformation are functionally indistinguishable from the solar-calibrated values for many objects. }
{ In this sense, the PHX-calibrated isochrones suggest a less severe issue with adopting the solar-calibrated mixing length for low-metallicity model stars.
We acknowledge that the PHX calibration yields more conservative results (in some cases; in others, the PHX values {demand} questionably large mixing lengths), but we emphasize again that this calibration scheme is less empirical than the VC transformation.}
The difference between fits found with PHX and VC alone suggest it would be worthwhile to compare color calibrations at low metallicities in a more systematic way.

Excluding HIP 54693---which displays the most extreme { deviation from $\alpha_{\odot}$}---the PHX-calibrated, best-fitting mixing lengths for M92 and the parallax stars rise to or exceed their solar-calibrated values.

{ To illustrate the difference between the two color calibrations,} Figure \ref{color} shows two isochrones with identical input physics (atmosphere, mixing length, diffusive efficiency), but employing different color calibrations, against both M92 (top) and the four metal-poor subdwarfs (bottom). 
As this and Tables \ref{bestfitVC} and \ref{bestfitPHX} demonstrate, in some cases, $\alpha_{\text{MLT, PHX}}$ is twice the value of $\alpha_{\text{MLT, VC}}$.

A caveat with this observation is that the impact of changing $\alpha_{\text{MLT}}$ does not scale linearly with $\alpha_{\text{MLT}}$; as mentioned previously, relatively larger changes to the mixing length are required to adjust the temperature of an isochrone for high mixing length values. This varies likewise with magnitude. For example, { the $T_{\rm eff}$ response to a change of $\alpha_{\text{MLT}}$ from 2.5 to 3.0 }
at HIP 46120's magnitude is only $\sim30$K, whereas the { $T_{\rm eff}$ response to a change in $\alpha_{\text{MLT}}$ from 0.5 to 0.7} at the same magnitude is roughly 90 K. For this reason, we should not consider $\alpha_{\text{MLT}}$ values exceeding 2.5, or $1.3 \alpha_{\odot}$,  to be as comparably severe as $0.7 \alpha_{\odot}$ ($\alpha_{\text{MLT}}=1.35$). { The decrease in $T_{\rm eff}$ sensitivity means that the impact of (1) uncertainty in observational constraints and (2)  degenerate parameter effects (e.g. $Z_{0}$-$\alpha_{\rm MLT}$ covariance) in the high-$\alpha_{\text{MLT}}$ regime are enhanced. Specifically, a comparably severe adjustment in the absolute value of $\alpha_{\rm MLT}$ would be necessary to exceed the the upper bound of a $T_{\rm eff}$ observational constraint, as compared to that required at the $T_{\rm eff}$ lower bound. The same change (uncertainty) in $Z_{0}$ mimics a larger $\Delta \alpha_{\rm MLT}$ for large $\alpha_{\rm MLT}$ than for small $\alpha_{\rm MLT}$. }

Regardless, the difference imparted by the color calibration alone speaks to the importance of HD 140283. Knowing this star's radius and luminosity allows us to infer its temperature from first principles, thereby providing a much more stringent test of the stellar models and eliminating this source of uncertainty entirely.

%-------------------------- last updated 7/24/17--------------------------------------
\begin{table} 
\centering 
\caption{Mixing Length Values of Best-Fitting Isochrones to Main Sequence Stars} 
\begin{tabular}{ l l  l l l l  l}  
\hline\hline
 { Object  }& {Model Atm} & {$\eta_{\text{D}}$} & $\FeH$ &  {$\alpha_{\text{MLT}}$} 
 & {$\alpha_{\text{MLT}}/\alpha_{\odot}$} & {Age }  \\ \hline

HIP46120   &Phx 			& 1.0		&-2.25	& 1.85	&0.9606			&12 \\		
			& grey	& 1.0		&-2.25	& 1.6 	&0.879		&13 \\
			&Phx 			& 0.5 	 	&-2.25	& 1.55	&0.847		&13 \\
			&Phx 			& 1.5 		&-2.25	& 1.97 	&0.996		&12  \\
\hline
HIP54639 	&Phx 			& 1.0		&-2.5	& 0.7	&0.363		&15 \\		
			& grey	& 1.0		&-2.5	& 0.5 	&0.275 		&15 \\
			&Phx 			& 0.5 	 	&-2.5	& 0.6	&0.328		&13 \\
			&Phx 			& 1.5 		&-2.5	& 0.7 	&0.353			&13  \\
\hline
HIP106924 	&Phx 			& 1.0		&-2.25	& 1.1	&0.571		&13 \\		
			& grey	& 1.0		&-2.25	& 0.95 	&0.522		&13 \\
			&Phx 			& 0.5 	 	&-2.25 & 1.0		&0.547			&13 \\
			&Phx 			& 1.5 		&-2.25 & 1.2 	&0.6067			&12  \\
\hline
Wolf1137 	&Phx 			& 1.0		&-2.5	& 1.95	&1.012	&15 \\		
			& grey	& 1.0		&-2.5	& 1.6 	&0.879	&14 \\
			&Phx 			& 0.5 	 	&-2.5	& 1.65	&0.902			&15 \\
			&Phx 			& 1.5 		&-2.5	& 2.1 	&1.062		&13  \\
\hline
\end{tabular}
\tablecomments{
The same data are presented as in Table \ref{M92only}, excluding distance modulus and reddening, for fits to each of the metal-poor subdwarfs. A VC color calibration is used for all isochrones tested.
}
\label{bestfitVC}
\end{table}

%-------------------------------------------------------------------------------------------------
\begin{figure} % figure: demonstrates configuration combinations that fit stars: REDO WITH PROPER TRACKS 	
\centering
\includegraphics[width=\linewidth]{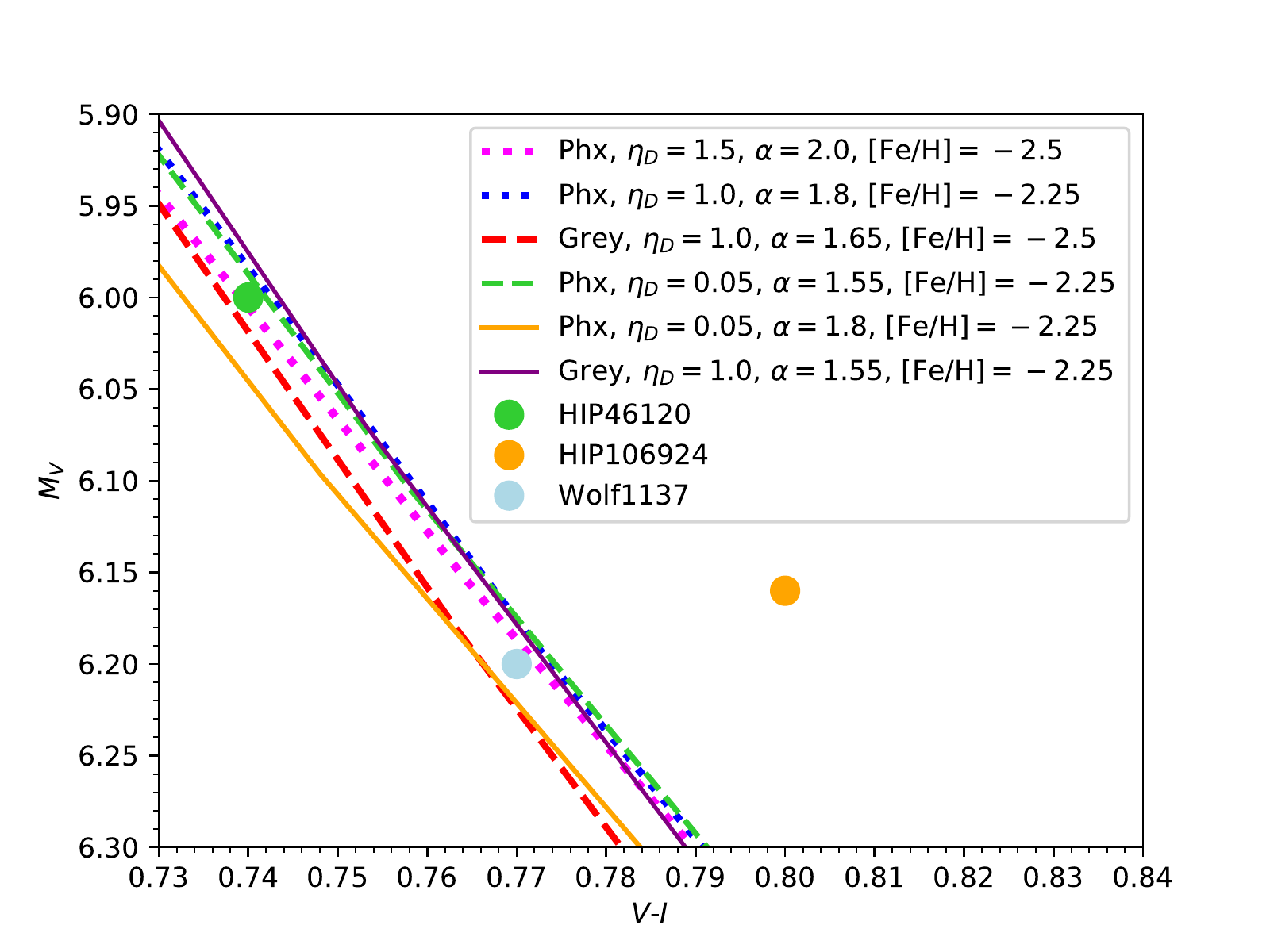}
\caption{
A sample of isochrones invoking input parameter combinations, metallicities, and mixing lengths tested against HIP 46120, HIP 106924, and Wolf 1137 is shown. All use a VC color calibration; details of the physical configuration are given in the legend.
}
\label{parfit}
\end{figure}
%-------------------------------------------------------------------------------------------------

%-------------------------------------------------------------------------------------------------
\begin{figure} % figure: demonstrates configuration combinations that fit stars: REDO WITH PROPER TRACKS 	
\centering
\includegraphics[width=\linewidth]{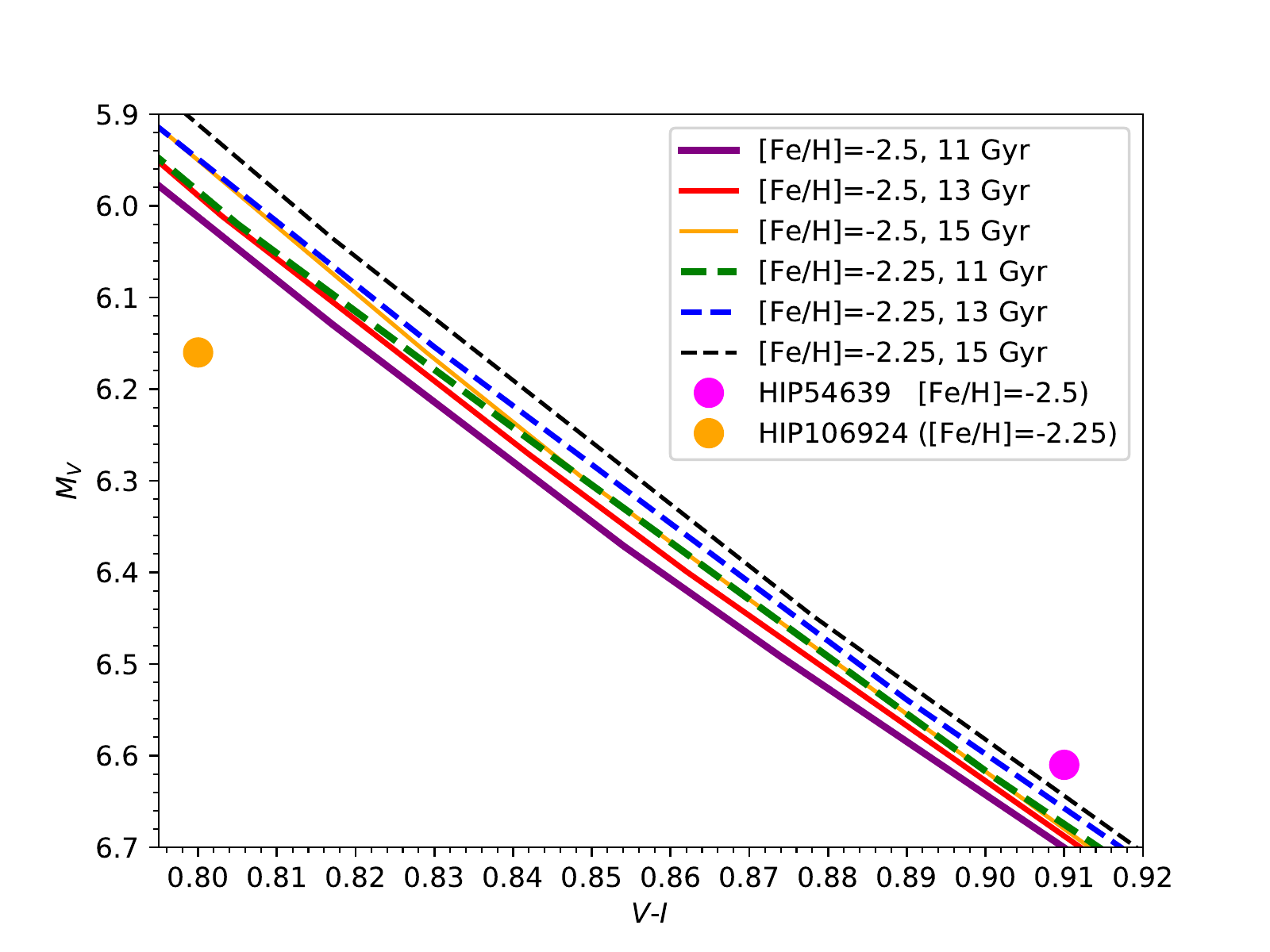}
\caption{
Isochrones with ages 11, 13, and 15 Gyr are shown at two metallicities: $\FeH=-2.25$ and $-2.5$. Both sets employ a PHOENIX model atmosphere,$\alpha_{\text{MLT}}=0.7$, $\eta_{\text{D}}=0.5$, and a VC color calibration. Subdwarfs HIP 54639 and HIP 106294---the two requiring the lowest fitted mixing lengths---are shown as pink and orange circles, respectively, for context.
}
\label{twoZ}
\end{figure}
%-------------------------------------------------------------------------------------------------

%-------------------------- last updated 7/24/17 ----------------------------------------------
\begin{table*} 
\centering 
\caption{Parameters of Best-Fits to M92 using a PHX Color Calibration }
\begin{tabular}{ l l  l l l l  l l l}  
\hline\hline
 { Object  }& {Model Atm} & {$\eta_{\text{D}}$} & $\FeH$
 & $m-M_V$ & E(B - V)   
 & {$\alpha_{\text{MLT}}$} & {$\alpha_{\text{MLT}}/\alpha_{\odot}$} & {Age }  \\ \hline
M92			&Phx 			& 1.0		&-2.4 & 14.7 & 0.05 & 1.9 	&0.986		&12 \\		
			&grey			& 1.0		&-2.4 & 14.7 & 0.06	& 1.8 	&0.988		&12 \\
			&Phx 			& 0.5 	 	&-2.4 & 14.7 & 0.06	& 1.9 	&1.039		&13 \\
			&Phx 			& 1.5 		&-2.4 & 14.7 & 0.05 & 1.97 	&0.996		&13  \\
\hline
\end{tabular}
\tablecomments{ 
The same data as given in Table \ref{M92only} is shown using isochrones calibrated instead to a PHX color scheme. 
}
\label{M92PHX}
\end{table*}

\begin{table} 
\centering 
\caption{Mixing Length Values of Best-Fitting Isochrones to Subdwarfs using a PHX Color Calibration }
\begin{tabular}{ l l  l l l l  l}  
\hline\hline
 { Object  }& {Model Atm} & {$\eta_{\text{D}}$} & $\FeH$  & {$\alpha_{\text{MLT}}$}
  & {$\alpha_{\text{MLT}}/\alpha_{\odot}$} & {Age }  \\ \hline
% M92			&Phx 			& 1.0		&-2.4	& 1.9 	&0.986		&12 \\		
% 			& grey	& 1.0		&-2.4	& 1.8 	&0.988		&12 \\
% 			&Phx 			& 0.5 	 	&-2.4	& 1.9 	&1.039		&13 \\
% 			&Phx 			& 1.5 		&-2.4	& 1.97 	&0.996		&13  \\
% \hline
HIP46120   &Phx 			& 1.0		&-2.25	& 3.0	&1.558		&13 \\		
			& grey	& 1.0		&-2.25	& 3.0 	&1.648		&13 \\
			&Phx 			& 0.5 	 	&-2.25	& 2.5	&1.367		&14 \\
			&Phx 			& 1.5 		&-2.25	& 1.3 	&1.517		&11  \\
\hline
HIP54639 	&Phx 			& 1.0		&-2.5	& 1.0	&0.519		&14 \\		
			& grey	& 1.0		&-2.5	& 0.8 	&0.439		&15 \\
			&Phx 			& 0.5 	 	&-2.5	& 0.9	&0.492		&13 \\
			&Phx 			& 1.5 		&-2.5	& 1.1 	&0.556		&14  \\
\hline
HIP106924 	&Phx 			& 1.0		&-2.25	& 1.75	&0.9087		&15 \\		
			& grey	& 1.0		&-2.25	& 1.44 	&0.791		&13 \\
			&Phx 			& 0.5 	 	&-2.25	& 1.44	&0.787		&13 \\
			&Phx 			& 1.5 		&-2.25	& 1.9 	&0.9606		&12  \\
\hline
Wolf1137 	&Phx 			& 1.0		&-2.5	& 3.0	&1.558		&11 \\		
			& grey	& 1.0		&-2.5	& 2.5 	&1.373		&11 \\
			&Phx 			& 0.5 	 	&-2.5	& 3.0	&1.640		&15 \\
			&Phx 			& 1.5 		&-2.5	& $>3.0$ &$>1.52$		&15  \\
\hline
\end{tabular}
\tablecomments{ 
The same data as given in Tables \ref{M92only} and \ref{bestfitVC} is shown using isochrones calibrated instead to a PHX color scheme. 
}
\label{bestfitPHX}
\end{table}

%--------------------------updated 6/24/17---------------------------- 
\begin{table*} 
\centering 
\caption{Summary: Best-Fitting Mixing Lengths to All Objects}
\begin{tabular}{ l l  l l l  l l  }  
\hline\hline
&  &  {Default }& {Average } &  &  &  \\
%\hline
 { Object  }& {Evolutionary Phase} &  { $\alpha_{\text{MLT}}$  }& { $\alpha_{\text{MLT}}/\alpha_{\odot}$}
 & { $\alpha_{\text{MLT}}/\alpha_{\odot}$ } & {Age (Gyr)} 	& Fit Method 
 \\ \hline
HD140283	& subgiant  			&1.3  	& 0.52	& 0.36-0.68	& 12.5 		& stellar track		\\		
M92			& Red Giant  			&1.75 	& 0.91	& 0.91	&13 		 	& isochrone	  	\\		
HIP46120    & main sequence         &1.85 	& 0.96 	& 0.92	&12 			& isochrone	 \\
HIP54639 	& main sequence  		&0.7  	& 0.36	& 0.33	&13 			& isochrone	 \\
HIP106924 	& main sequence  		&1.1    & 0.57 	& 0.56	&13 			& isochrone	 \\
Wolf1137 	& main sequence   		&1.95   & 1.01 	& 0.96	&12 			& isochrone	  \\
\hline
\end{tabular}
\tablecomments{
The object and evolutionary phase it represents are given in columns 1 and 2, respectively. Columns 3 and 4 give the best-fitting mixing length and normalized mixing length values for that object at the default configuration: PHOENIX model atmosphere,  $\eta_{\text{D}}=1.0$, and with a VC color calibration applied. Column 5 provides the average value of the normalized mixing length across the four physical prescriptions considered for M92 and the four subdwarfs. 
}
\label{bestfit}
\end{table*}
%\vskip5ex

%-------------------------------------------------------------------------------------------------
\begin{figure} % impact of color calibration 7/24/17
\centering
\includegraphics[width=\linewidth]{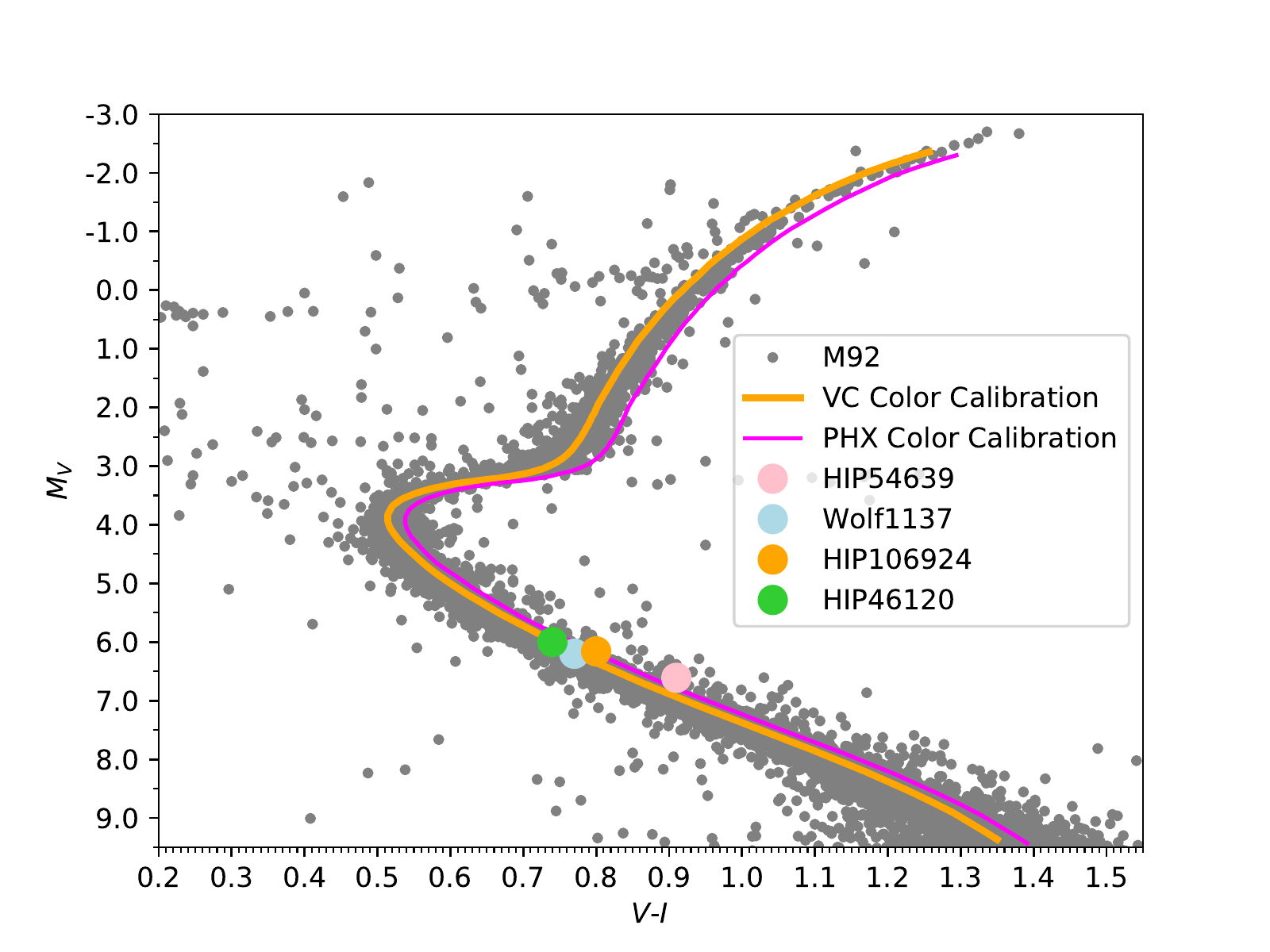}
\includegraphics[width=\linewidth]{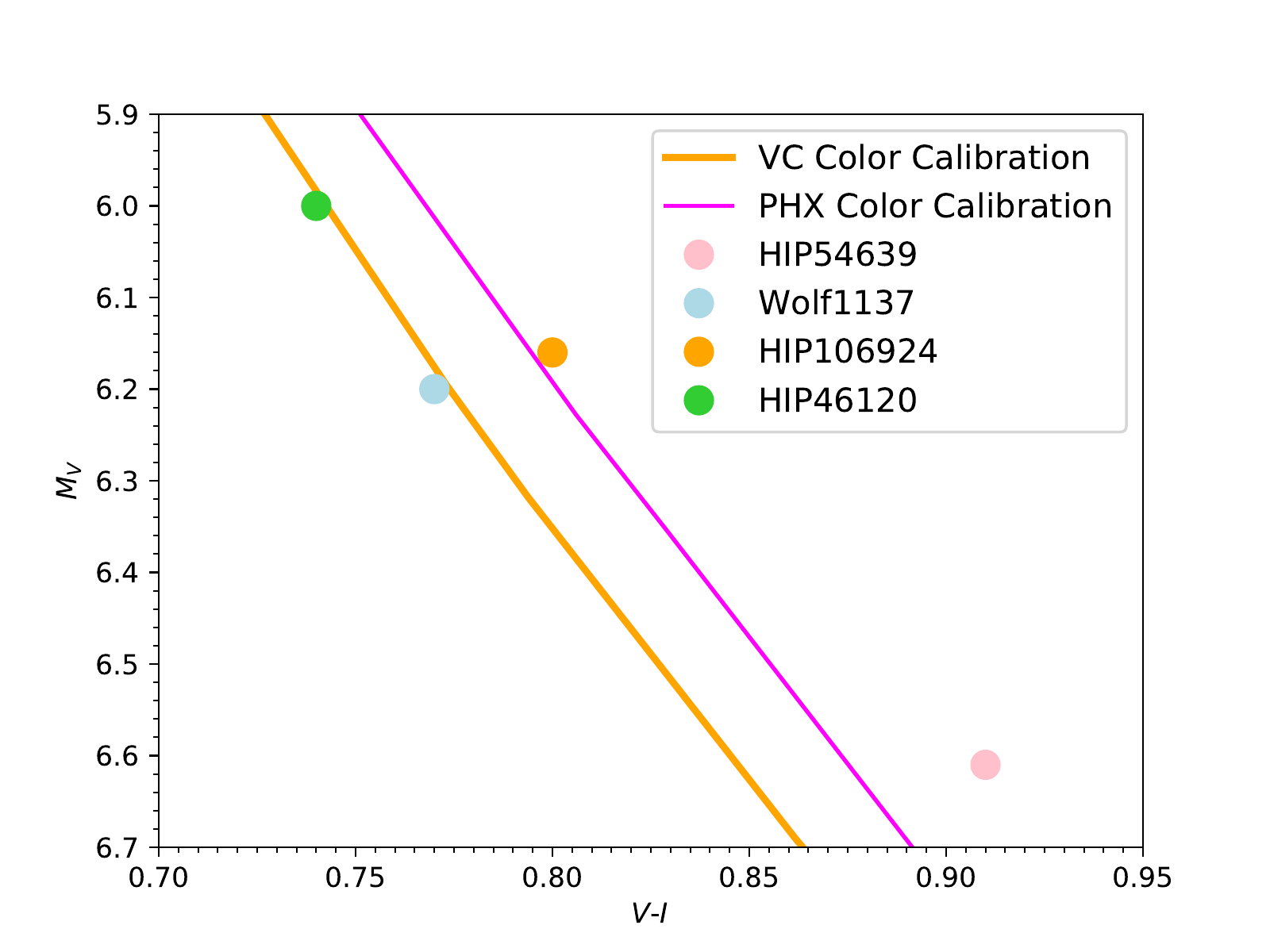}
\caption{TOP: Two isochrones with identical physical configurations (PHOENIX model atmosphere, $\eta_{\text{D}}=1.0$, $\alpha_{\text{MLT}}=1.8$) but using different color calibrations are shown against the M92 data at a reddening of 0.5 and distance modulus of 14.7. Age = 13 Gyr. BOTTOM: The same two color calibrations are shown against the four metal-poor parallax stars; e.g.\ the main sequence region. The physical configuration of the isochrones is the same as in Figure \ref{color}.
}
\label{color}
\end{figure}
%-------------------------------------------------------------------------------------------------
 
\section{Summary and Conclusions}

%%%%%%%%%%%%%%%%%%%%%%%%%%%%%%%%%%%%%%%%%%%%%%%%%%%%%%5

In 2015, Creevey et al.\ demonstrated that stellar models with a solar-valued mixing length could not reproduce observations of HD 140283, adding to a body of accumulating evidence suggesting that {the mixing length is not a universal constant and should not be treated as such.} 
In agreement with C15,  we also find that we cannot reproduce the observable features of HD 140283  with DSEP using a solar mixing length. To account for uncertainties in stellar models, we present the mass and mixing length combinations that reproduce HD 140283  under a range of physical input configurations; in particular, variations in atmospheric boundary conditions and the efficiency of diffusion. 
To contextualize our results in terms of the solar mixing length, we calibrate the mixing length within DSEP to yield solar specifications for each physical configuration. 
In all configurations, we require subsolar mixing lengths to reproduce observations of HD 140283.

{ To form a more well-rounded picture of the potential importance of non-solar $\alpha_{\rm MLT}$ values on metal-poor stars as a whole,}
we take other evolutionary phases into consideration. We generate sets of isochrones over a range of mixing length values to find the best possible fit to globular cluster M92, whose metallicity and age closely match that of HD 140283 { (where HD 140283's age is inferred from our stellar model fits)}. In addition, we examine fits of these isochrones to four highly metal-poor subdwarfs on the main sequence.

Table \ref{bestfit} summarizes our findings about the best-fitting mixing lengths for all objects.
The default mixing length is considered to be the value of $\alpha_{\text{MLT}}$ found to best-fit an object under a PHOENIX model atmosphere, $\eta_{\text{D}}=1.0$, and using a VC color calibration. The average normalized mixing length refers to the average of all mixing lengths as a proportion of their solar-calibrated values; e.g.\ the sum of the best-fitting PHOENIX, grey, $\eta_{\text{D}}=0.5$ and $\eta_{\text{D}}=1.5$ mixing length values per object, divided by four. In short, subsolar mixing lengths are found to provide the best reproductions of observations across the board.

Our results concerning the fit to HD 140283 are in strong agreement with the findings of C15, with DSEP's best-fitting mixing length falling between $0.42$ to $0.68$\% of DSEP's $\alpha_{\odot}$. If a strict age limit of 13 Gyr \citep{Planck} is imposed,  $\alpha_{\text{MLT}}/\alpha_{\odot}$ centers closer to $50$\%---precisely what the CESAM code required to fit HD 140283. Given the sensitivity of the subgiant branch to the mixing length parameter, and that HD 140283's interferometrically determined radius allows us to derive temperature constraints from first principles, this result should be taken with a degree of confidence beyond that which fits to other regimes can afford.

The best-fitting mixing length values found through isochrone fits to M92, in contrast, are both the highest-valued among any object we consider and subject to the broadest array of uncertainties. Our best fits employ mixing lengths at $\sim90$\% of the solar value, across all physical conditions. The length of the subgiant branch and curvature of the red giant branch are { sensitive to both $\alpha_{\rm MLT}$ and age---to the extent that for a given $\alpha_{\rm MLT}$, the age can be determined within 1 Gyr.} This finding is corroborated in part by \citet{Tayar}, who demonstrate that a variable mixing length must be invoked in order to correct a trend between temperature offset and metallicity among red giants. 
As in our case, \citet{Tayar}'s corrections demand smaller mixing lengths with decreasing metallicity, with their results suggesting that this could lead to errors approaching a factor of two in age determinations. This motivates continued efforts to understand the scope of validity of solar-valued $\alpha_{\rm MLT}$ and its empirical characterization in non-solar stellar interiors, {including understanding the behavior of $\alpha_{\rm MLT}$ with atmospheric parameters.}

{ We find, on average, that mixing lengths shorter than their solar-calibrated values are likewise required to fit the four main sequence parallax stars when the isochrones are calibrated according to a semi-empirical (VC) color transformation. We find no justification to claim an obvious empirical trend regarding mixing length along the main sequence---neither as a function of metallicity, nor concerning the relative best-fitting values in the main sequence stage compared to our findings in the other evolutionary phases---given the high variance in best-fitting $\alpha_{\text{MLT}}$ among these stars (and the fact that we have only four data points). However, the broad range of values ($\alpha_{\text{MLT}}=0.33$ to $0.96$) alone suggests that a critical re-evaluation is in order. }

When a theoretical color calibration is implemented instead of the VC calibration, best-fitting mixing lengths rise considerably. Mixing lengths near or even slightly above the solar value (per case) are found to fit best when isochrones adopted a transformation based on synthetic colors from Phoenix atmospheric models. While \citet{Tayar}, \citet{Creevey}, and recent findings from asteroseismology (e.g.\ \citet{Creevey17}) all support the need for subsolar mixing lengths, the best-fitting mixing lengths we found using the PHX calibration are in better agreement with the findings of \citet{Ludwig99}, \citet{Trampedach}, and \citet{MagicMLT}. {We note, however, that the work of \citet{Ludwig99} and \citet{Trampedach} is not necessarily generalizable to stars with the metal depletion considered here.} A detailed study of the effect of color calibrations, and the uncertainties therein, on mixing length calibrations specifically and on very metal-poor stars in general would be highly informative in this regard.

Despite this, and despite ambiguous results on the main sequence regarding particular values of $\alpha_{\text{MLT}}$, the subsolar consistency of our results across diverse evolutionary phases and physical conditions point collectively and conclusively to one key point: In order for stellar modeling to maintain fidelity, $\alpha_{\text{MLT}}$ must be treated as an adaptive free parameter, not a physical constant. 
We mean this in two capacities: 
(1) that $\alpha_{\text{MLT}}$ should be adjusted to a subsolar value when modeling very metal-poor stars (probably as a function of that star's metallicity, though we have intentionally considered only one [Fe/H] regime), and 
(2) that $\alpha_{\text{MLT}}$ should be adaptive { for a given mass}
according to evolutionary phase. 
{Because we have carefully chosen a sample of stars with very similar metallicities,}
our results support the notion that the same value of $\alpha_{\text{MLT}}$ should not be preserved across the main sequence, subgiant, and red giant phases {for a given mass}.

We have demonstrated that subsolar values of the mixing length provide better fits to highly metal-poor stellar objects, even when uncertainties in atmospheric boundary conditions, diffusion, metallicity, and color-calibration---to an extent---are taken into account. This finding is consistent with recent literature on { mixing length vs. metallicity trends,} and gains particular urgency in the era of asteroseismic observing capabilities. 

Highly accurate stellar models are necessary, among many things, for the accurate dating of globular clusters and for asteroseismic studies of the stellar systems hosting exoplanets. The use of an adaptive mixing length is undeniably crucial to the future of stellar modeling, but in this sense, it may well lead to important cosmological and observational insights as well.

\section{Acknowledgements}
This work is supported by grant AST-1211384 from the National Science Foundation. 

We would like to thank the referee for their insightful and detailed reports, and especially for their contribution to the discussion of the adiabat and the relationship between the size of a star's convective envelope and the size of the star as a whole. The referee has provided critical analysis, especially regarding section 6.2, which we have included in the text.

We would like to thank the University of Cape Town and the South African Astronomical Observatory for providing accommodation and laboratory resources for part of the time during which this study was being conducted. 

\bibliographystyle{apj}
\bibliography{complicatedbib_Dec17}

\end{document}